\documentclass[journal ]{new-aiaa}
\usepackage[utf8]{inputenc}
\usepackage{textcomp}

\usepackage{graphicx}
\usepackage{amsmath}
\usepackage[version=4]{mhchem}
\usepackage{siunitx}
\usepackage{lettrine}
\usepackage{url}
\usepackage{longtable}
\usepackage{tabularx}
\usepackage{booktabs}
\usepackage{subfigure}
\usepackage{bm}
\usepackage{color}
\usepackage{trimclip}
\usepackage{textcomp}

\DeclareMathOperator{\sign}{sgn}
\DeclareMathOperator{\derivd}{d}
\def\AR{\text{\itshape\clipbox{0pt 0pt .32em 0pt}\AE\kern-.30emR}}

\title{Applying frequency-domain unsteady lifting-line theory to time-domain problems}

\author{Hugh J. A. Bird \footnote{Graduate Researcher, h.bird.1@research.gla.ac.uk, Student Member AIAA}}
\author{Kiran Ramesh\footnote{Lecturer, Aerospace Science Division, School of Engineering, Kiran.Ramesh@glasgow.ac.uk, Member AIAA}}
\affil{Aerospace Sciences Division, School of Engineering, University of Glasgow, Glasgow, United Kingdom, G12 8QQ}

\begin{document}

\maketitle

\begin{abstract}
Frequency-domain unsteady lifting-line theory is better developed than its time-domain
counterpart. To take advantage  of this, this paper transforms time-domain 
kinematics to the frequency domain, performs a  convolution and then returns 
the results back to the time-domain. It demonstrates how well-developed frequency-domain
methods can be easily applied to time-domain problems, enabling prediction of forces and 
moments on finite wings undergoing arbitrary kinematics.
\end{abstract}

\section*{Nomenclature}

{\renewcommand\arraystretch{1.0}
\noindent\begin{longtable*}{@{}l @{\quad=\quad} l@{}}
$\AR$ & aspect ratio\\
$c$ & chord\\
$\overline{c}$ & mean chord\\
$C_l$ & two-dimensional lift coefficient\\
$C_L$ & three-dimensional lift coefficient\\
$C_m$ & two-dimensional moment coefficient\\
$C_M$ & three-dimensional moment coefficient\\
$F$ & three-dimensional correction strength\\
$g$ & return ramp function\\
$h$ & heave displacement\\
$h_0^*$ & non-dimensional heave oscillation amplitude\\
$k$ & chord reduced frequency\\
$K$ & three-dimensional interaction kernel\\
$P$ & ramp-hold-return amplitude\\
$s$ & semispan\\
$t$ & time\\
$U_\infty$ & freestream velocity\\
$w_0,w_1,w_2$ & quadratic interpolation weights\\
$x,y,z$ & coordinate system with origin at wing root\\
$x_m^*$ & non-dimensional reference location for pitching moment\\
$x_p^*$ & non-dimensional pivot location for pitching\\
$\alpha$ & pitch angle\\
$\alpha_0$ & pitch oscillation amplitude\\
$\Gamma$ & bound circulation\\
$\zeta$ & spanwise coordinate\\
$\nu$ & span reduced frequency\\
$\sigma$ & non-dimensional smoothing parameter for ramp-hold-return\\
$\phi$ & velocity potential\\
$\omega$ & angular frequency\\
\end{longtable*}}

\section{Introduction}

Lifting-line theory provides an elegant means by which to 
apply two-dimensional (2D) aerodynamic models to three-dimensional (3D) finite wing problems \cite{Prandtl1923, Dyke1964}.
 By correcting 2D solutions
for 3D effects, lifting-line theory avoids the complications
of fully 3D solutions. Consequently, it can reduce computational
cost and lead to a better understanding of the
problem being studied. 

The need for low-order methods to model the unsteady aerodynamics
of finite wings in unsteady flow has lead to the development of 
Unsteady Lifting-Line Theory (ULLT). Compared to strip theory \cite{Leishman2006}, which simply sums the 2D solutions for the chord distribution along the wing,
it doesn't neglect the often important finite-wing effects. The computational
cost is far lower than fully 3D numerical methods such as the unsteady vortex lattice method \cite{Simpson2016},
vortex particle methods \cite{Bird2021, Willis2007, Winckelmans1993}, or computational fluid dynamics which often requires
high performance computing resources. This is important in a world where the unsteady
aerodynamic problems involving finite wings are becoming increasingly common.

Large scale problems involve high altitude long endurance aircraft~\cite{Shearer2007} and ever-larger wind turbines~\cite{Hansen2006}. 
As scale increases, increased flexibility in the wing or turbine blade means that studying the 
unsteady aerodynamic response becomes more important. More pedestrian unmanned aerial vehicles
also encounter challenges due to unsteady gusts at low altitude~\cite{Williams2002,Jones2020}. At smaller scales, a better
understanding of unsteady aerodynamics eases the design of both oscillating energy harvesting 
devices~\cite{Rostami2017} and also micro air vehicles that mimic insects using flapping wing technology.

Frequency-domain unsteady lifting-line theory is well developed. It is in effect a 3D
extension of the early frequency-domain work of Theodorsen \cite{Theodorsen1935} and Sears \cite{Sears1941}. 
Lifting-line theory is based on the assumption that the 
chord is much smaller than the span. When looking at the
detailed flow around the chord, the span is so big as to be unimportant -
the flow can be assumed to be 2D. When looking at the span-scale 3D problem involving
the finite wing, the chord is sufficiently small that its influence can be assumed
to just be on a (lifting-)line. Neither of these 2D or 3D models have enough
information to model the complete 3D flow on their own. Lifting-line theory links them together
to obtain a solution. The 3D model uses the bound circulation obtained from the 2D model, and
the 2D model accounts for 3D effects using a correction obtained from the 3D model.

Work on frequency-domain unsteady lifting-line theory by 
James \cite{James1975}, Van Holten \cite{Holten1976}, Cheng \cite{Cheng1976},
Ahmadi and Widnall \cite{Ahmadi1985} and Sclavounos \cite{Sclavounos1987}
improved on the wake model by which the 2D inner problem was corrected by the simplified 3D outer
problem. The importance of this wake model was demonstrated by Bird and Ramesh~\cite{Bird2021b}.
These models were followed by Guermond and Sellier's unsteady lifting-line theory valid for swept and 
curved wings oscillating at all frequencies \cite{Guermond1991}.

Famously, Wagner's \cite{Wagner1925} time-domain equivalent to Theodorsen's
problem \cite{Theodorsen1935, Garrick1938} is challenging to evaluate exactly. A consequence of 
this is that common approximations of Wagner's function are based on 
the inverse Laplace transform of Theodorsen's function \cite{Fung1993, Dowell1980,Dawson2021}.
This inverse Laplace transform depends upon the linearity of the underlying theory.
The same property was used by \=Otomo et al.~\cite{Otomo2020} to apply Theodorsen to
periodic but non-sinusoidal kinematics.
Analytical time-domain methods are arguably inherently more challenging
to formulate than their frequency-domain counterparts. 
It is therefore unsurprising that progress in
time-domain unsteady lifting-line theory has been less rapid
than the frequency-domain counterpart. 

The earliest time domain method is that of Jones \cite{Jones1940}, where
a simplified solution for an elliptic wing was obtained. More recently, Boutet and Dimitriadis \cite{Boutet2018}
combined a Prandtl-like wake with Wagner's theory. However, as was found by Bird and Ramesh \cite{Bird2021b},
Prandtl-like wake models obtain different solutions in comparison with more complete wake models such as those used by 
Guermond and Sellier \cite{Guermond1991}.

Numerical methods have enabled more complete wake models in the time-domain.
Phlips et al. \cite{Phlips1981} and Nabawy and Crowther \cite{Nabawy2015} both constructed numerical methods for flapping flight
using Prandtl-like wake models.
Devinant \cite{Devinant1998} constructed a numerical time-domain method based upon a simplified
version of Guermond and Sellier's frequency-domain theory \cite{Guermond1991}.
This time marching theory required straight wings and uniform 3D induced downwash,
in theory limiting its applicability to only low-frequency kinematics. However, Bird and Ramesh \cite{Bird2021b}
found that this theoretical limitation had limited practical relevance for rectangular wing planforms.
Crucially though, it modeled spanwise vorticity in the wake of the inner 2D domain and the correction
for the change of the wake's spanwise vorticity with respect to span in the outer 3D domain.
This method was extended to large amplitude kinematics by Bird et al. \cite{Bird2019} using 
Ramesh et al.'s \cite{Ramesh2013} discrete vortex method for the 2D solutions.
This 2D solution was also used by Ramesh et al. \cite{Ramesh2017} in combination with a
Prandtl-like wake. 

In the time domain, analytical methods for finite wings are currently confined to simplified wake
models. This limits their accuracy \cite{Bird2021b}. More complicated wake models can
be constructed using numerical methods, but numerical methods
typically entail increased computational costs and convergence issues.
As a consequence, in practice strip theory is often used for the 
prediction of finite-wings undergoing unsteady motion.
This paper aims to circumvent the challenge of formulating an analytical
time-domain unsteady lifting-line theory by instead applying a frequency-domain 
ULLT to time-domain problems using Fourier transforms to perform the convolution of 
the input kinematics with the wing's response in the frequency domain, taking
advantage of the linearity of analytical frequency-domain ULLTs.
To avoid the evaluation of the frequency-domain method at every frequency,
a quadratic interpolation scheme is used. The interpolation requires only a small
number of evaluations of the frequency-domain theory to obtain good results, and
need only be evaluated once for a given geometry, regardless of kinematics.
In this paper, this ULLT / Convolution in Frequency Domain (UCoFD) method is 
applied to Sclavounos' ULLT~\cite{Sclavounos1987}. In theory however,
the approach is independent of the details of the underlying ULLT.

This paper is laid out as follows: first, the frequency-domain ULLT used in this paper is
detailed in Sec.~\ref{sec:freq_domain}. This ULLT follows the work of Sclavounos with
minor modifications. Next, the interpolation of frequency-domain results and
the method to apply frequency-domain results to time-domain problems is introduced
in Sec.~\ref{sec:td_to_fd}. UCoFD is then compared to CFD and strip theory for
both inviscid problems and low Reynolds
number problems, relevant to modern applications, in Sec.~\ref{sec:results}. Concluding remarks are made in
Sec.~\ref{sec:conclusions}.

\section{Frequency-domain lifting-line theory}
\label{sec:freq_domain}

An unsteady lifting-line theory, based on the work of Sclavounos~\cite{Sclavounos1987}, is presented in this section.
Pitch and heave oscillations are considered. A more detailed derivation is given in Bird and Ramesh~\cite{Bird2021b}.

Sclavounos's ULLT considers a wing in inviscid, incompressible flow
undergoing small-amplitude oscillation. The freestream velocity of the 
flow is $U_\infty$, and the heave $h$ and pitch $\alpha$ displacement of the wing respectively are 

\begin{equation}
 	h(y;t) = h_0e^{i \omega t} = h^*_0(y)c(y)e^{i \omega t},
	\label{eqn:kinem_heave}
\end{equation}
\begin{equation}
 	\alpha(y;t) =\alpha_0e^{i \omega t},
	\label{eqn:kinem_pitch}
\end{equation}

\noindent where $t$ is time, $c$ is chord and $\omega$ is the angular frequency
of oscillation. 
$h_0^*$ is the heave oscillation amplitude $h_0$ non-dimensionalized by the chord, $\alpha_0$
is the pitch oscillation amplitude and $y$ is a coordinate on the span of the wing.
The oscillation frequency can be non-dimensionalized
by either the chord or span as

\begin{equation}
	 k(y) = \frac{\omega c(y)}{2U_\infty},
\end{equation}
\begin{equation}
	 \nu = \frac{\omega s}{U_\infty},
\end{equation}

\noindent where $k(y)$ is the local chord reduced frequency and
$\nu$ is the span reduced frequency, non-dimensionalized by $s$, the semi-span of the wing.
For the rectangular wings studied in this paper, the chord is constant with respect
to span meaning $k(y)$ is constant. Consequently, it referred to as the 
chord reduced frequency for hereon. 

Lifting-line theory assumes high aspect ratio. 
The inner solution is obtained on the chord scale, and can be modeled in 2D since
it changes slowly with respect to span. It allows a detailed solution of flow around
the chord from which quantities such as lift, moment and bound circulation
can be obtained.
The outer solution is obtained on the span-scale and is 3D. The chord-scale
is negligible compared to the span-scale, so instead of the solving the difficult problem
of 3D surface-wake interaction, the simpler problem of interaction between a (lifting-)line and 
the wake can be considered.
These so-called inner and outer problems
interact. The outer 3D problem needs detail from the inner 2D problem
to obtain the bound circulation distribution, and 2D inner problem
requires a correction from the outer 3D problem to account for
the 3D nature of the flow. This leads
to the velocity potential being taken as a 2D solution with a 3D correction
is given by

\begin{equation}
 	\phi(x,y,z;t) \approx \phi^{2D} + F(y)(i\omega z e^{i \omega t} - \phi^{2D}_{hn}),
 	\label{eq:phi_llt}
\end{equation}

\noindent where $\phi^{2D}$ represents the velocity potential solution to the 2D Theodorsen problem and
the second term represents the correction to include 3D effects. 
This correction is of complex amplitude $F(y)$ and includes an oscillating uniform unit downwash velocity
potential $i \omega z e^{i \omega t}$ and the reaction of the 2D airfoil section
to this, $\phi_{hn}^{2D}$. 
$\phi_{hn}^{2D}$ is the 2D velocity potential solution of a section oscillating in heave
with unit amplitude.

Theodorsen's method can be used to obtain the 2D bound circulation  $\Gamma^{2D}$ 
associated with $\phi^{2D}$ for both pitch and heave:
\begin{subequations}
	\label{eq:bound_vort_2D}
\begin{align}
	\Gamma^{2D}_{h}(y;t) =& \frac{4 U_\infty h^*_0(y)c(y) e^{-ik}}{i H_0^{(2)}(k) + H_1^{(2)}(k)}e^{i \omega t} ,\\
	\Gamma^{2D}_{\alpha}(y;t) =& \frac{4U_\infty\alpha_0(y)c(y) e^{-ik}}{i H_0^{(2)}(k) + H_1^{(2)}(k)}\left(\left(x_p^*-\frac{3}{4}\right) -  \frac{1}{2 i k}\right)e^{i (\omega t)},
\end{align}
\end{subequations}
\noindent where $H_0^{(2)}(z)$ and $H_1^{(2)}(z)$ are Hankel 
functions of the second kind~\cite{Olver2010} and $x_p^*$ is the non-dimensional pivot location for pitching, where
$x_p^*=0$ indicates the leading edge and $x^*_p=1$ indicates the trailing edge.
The subscripts $h$ and $\alpha$ indicate the solutions to the heave problem and the
pitch problems respectively.

Following Eq.~\ref{eq:phi_llt}, this allows the corrected bound circulation to be obtained as
\begin{equation}
	\Gamma(y;t) = \Gamma^{2D}(y;t) - F(y)\Gamma^{2D}_{hn}(t) ,
	\label{eq:bound_vort_w_3D_correction}
\end{equation}
where subscript $hn$ indicates the bound circulation solution for the unit heave amplitude
$h_0=1$.

The strength of this 3D correction, $F(y)$, can be obtained from a known 
bound circulation distribution.
For frequency-domain ULLT, Sclavounos gives this as
\begin{equation}
	F(y) = -\frac{1}{2 \pi i \omega e^{i \omega t}} \int_{-s}^s \Gamma'(\eta)K(y - \eta)\derivd \eta,
        \label{eqn:match2}
\end{equation}
\noindent where $\Gamma'$ indicates the derivative of the bound circulation with respect to the spanwise coordinate $y$, and
\begin{equation}
	\label{eq: K unsteady}
	K(y) = \frac{1}{2s}\sign(y^*)\left[\frac{e^{-\nu |y^*|}}{|y^*|} - i\nu E_1(\nu|y^*|) + \nu P(\nu|y^*|) \right].
\end{equation}
$E_1(x)$ is the exponential integral~\cite{Olver2010}, $y^*=y/s$ is the normalized span coordinate and 
\begin{equation}
 	P(y) = \int^\infty_1 e^{-yt}\left[ \frac{\sqrt{t^2 - 1} - t}{t}\right]\derivd t  + i \int^1_0 e^{-yt}\left[\frac{\sqrt{1-t^2}-1}{t}\right] \derivd t.
\end{equation}
The interaction kernel $K(y)$ represents the correction for the difference between the inner 2D wake model and outer 3D wake model. 
In strip theory, where there is no 3D interaction, this is set to zero, giving $F_{\text{strip}}=0$.
Alternate, simplified interaction kernels are considered in Bird et al.~\cite{Bird2021b}.

$F(y)$ can be substituted into Eq.~\ref{eq:bound_vort_w_3D_correction} to obtain a differential equation
that allows the inner and outer solution to be solved simultaneously:
\begin{equation}
  \label{eq: integro-diff equation}
  \Gamma (y,t) - \frac{\Gamma^{2D}_{hn}(t)}{2 \pi i \omega} \int^s_{-s} \Gamma'(\eta)K(y-\eta) \derivd\eta = \Gamma^{2D}(y,t),
\end{equation}
\noindent Usually, a solution is obtained by approximating the bound circulation distribution as a
truncated Fourier series
\begin{equation}
	\Gamma(y, t) = 4U_\infty s \sum_{m=1}^M \Gamma_{m} \sin (m\zeta) e^{i\omega t},
\end{equation}
\noindent where $y = -s \cos(\zeta)$. This is solved at collocation points distributed over the span. 

Theodorsen's method allows the 2D lift and moment coefficients $C_l$ and $C_m$ to obtained for both
pitch and heave problems
\begin{subequations}
\label{eq:cl_cm}
\begin{align}
C_{l_h}(y;t) &= 2\pi h^*_0 (-2ikC(k) + k^2 ) e^{i \omega t} \label{eq:2d_C_l_h},\\
 C_{l_{\alpha}}(y;t) &= 2\pi\alpha_0\left[C(k)\left(1-2ik\left(x^*_p-\frac{3}{4}\right)\right)+\frac{ik}{2}+k^2\left(x^*_p-\frac{1}{2}\right)\right]e^{i (\omega t + \psi)}, \\ 
 C_{m_h}(y;t) &= 2\pi h^*_0 \left[ - 2ik C(k) \left(x^*_m-\frac{1}{4}\right) + k^2\left(x^*_m-\frac{1}{2}\right)\right] e^{i \omega t}   \label{eq:2d_C_m_h}, \\
 C_{m_{\alpha}}(y;t) &= 2\pi\alpha_0\Bigg[C(k)\left(1-2ik\left(x^*_p-\frac{3}{4}\right)\right)\left(x^*_m-\frac{1}{4}\right) \nonumber \\
       &+ k^2\left(x^*_p\left(x^*_m-\frac{1}{2}\right) - \frac{1}{2}\left(x^*_m-\frac{9}{16}\right)\right)+\frac{ik}{2}\left(x^*_m-\frac{3}{4}\right)\Bigg] e^{i (\omega t + \psi)},
 \end{align}
\end{subequations}
\noindent where $C(k) = \frac{K_1(ik)}{K_1(ik) + K_0(ik)}$ is Theodorsen's 
function in terms of modified Bessel functions of the second kind~\cite{Olver2010},
and $x_m^*$ is the non-dimensional reference location about which the pitching moment is calculated.

These force coefficients can be corrected for 3D effects using Eq.~\ref{eq:phi_llt} as
\begin{subequations}
\begin{align}
	C_{l}(y;t) &= C_{l}^{2D}(y;t) - F(y)C_{l,hn}^{2D}(y;t) ,\\
	C_{m}(y;t) &= C_{m}^{2D}(y;t) - F(y)C_{m,hn}^{2D}(y;t) .
\end{align}
\end{subequations}
where $C_{l,hn}^{2D}$ and $C_{m,hn}^{2D}$ indicated the lift and moment
coefficient for an airfoil oscillating in heave with amplitude $h_0=1$.

For a whole wing with mean chord $\overline{c}$, this allow lift and moment coefficients to be found as
\begin{equation}
  C_L = \frac{1}{2s \overline{c}}\int_{-s}^{s} C_l(y)c(y)\derivd y, \qquad C_M = \frac{1}{2s \overline{c}^2}\int_{-s}^{s} C_m(y)c^2(y)\derivd y .
\end{equation}

\section{Applying frequency-domain methods to time-domain results}
\label{sec:td_to_fd}

For the UCoFD method to apply frequency-domain ULLT to time-domain problems, 
frequency domain solutions must first be obtained for the entire frequency spectrum. 
Evaluating the frequency-domain solution at every frequency is
computationally expensive and unnecessary. Instead, the solutions can be interpolated with
respect to frequency.

A lifting-line theory takes a 2D method and corrects for 3D effects. The 2D inner solution is dominant. 
Examining the lift and moments obtained from 2D (Eq.~\ref{eq:cl_cm}), it can be observed they are of $O(k^2)$. They 
can be interpolated quadratically using the weighting scheme
\begin{subequations}
\begin{align}
	w_0(\omega) &= \frac{(\omega - \omega_1)(\omega - \omega_2)}{(\omega_0 - \omega_1)(\omega_0 - \omega_2)},\\
	w_1(\omega) &= \frac{(\omega - \omega_0)(\omega - \omega_2)}{(\omega_1 - \omega_0)(\omega_1 - \omega_2)},\\
	w_2(\omega) &= \frac{(\omega - \omega_0)(\omega - \omega_1)}{(\omega_2 - \omega_0)(\omega_2 - \omega_1)},
\end{align}
\end{subequations}
which can then be applied to obtain $C_L$ or $C_M$ as
\begin{equation}
	C_{L/M}(\omega) = w_0(\omega)C_{L/M}(\omega_0) + w_1(\omega)C_{L/M}(\omega_1) + w_2(\omega)C_{L/M}(\omega_2)
\end{equation}
where $\omega_0 < \omega_1 < \omega_2$ and $\omega_0 < \omega < \omega_2$ when interpolating. Extrapolation 
is necessary at high frequencies.

\begin{figure}
\centering
    \subfigure[Heave $C_L$]{
	\label{fig:heave_CL_interp}
    	\includegraphics[width=0.4\textwidth]{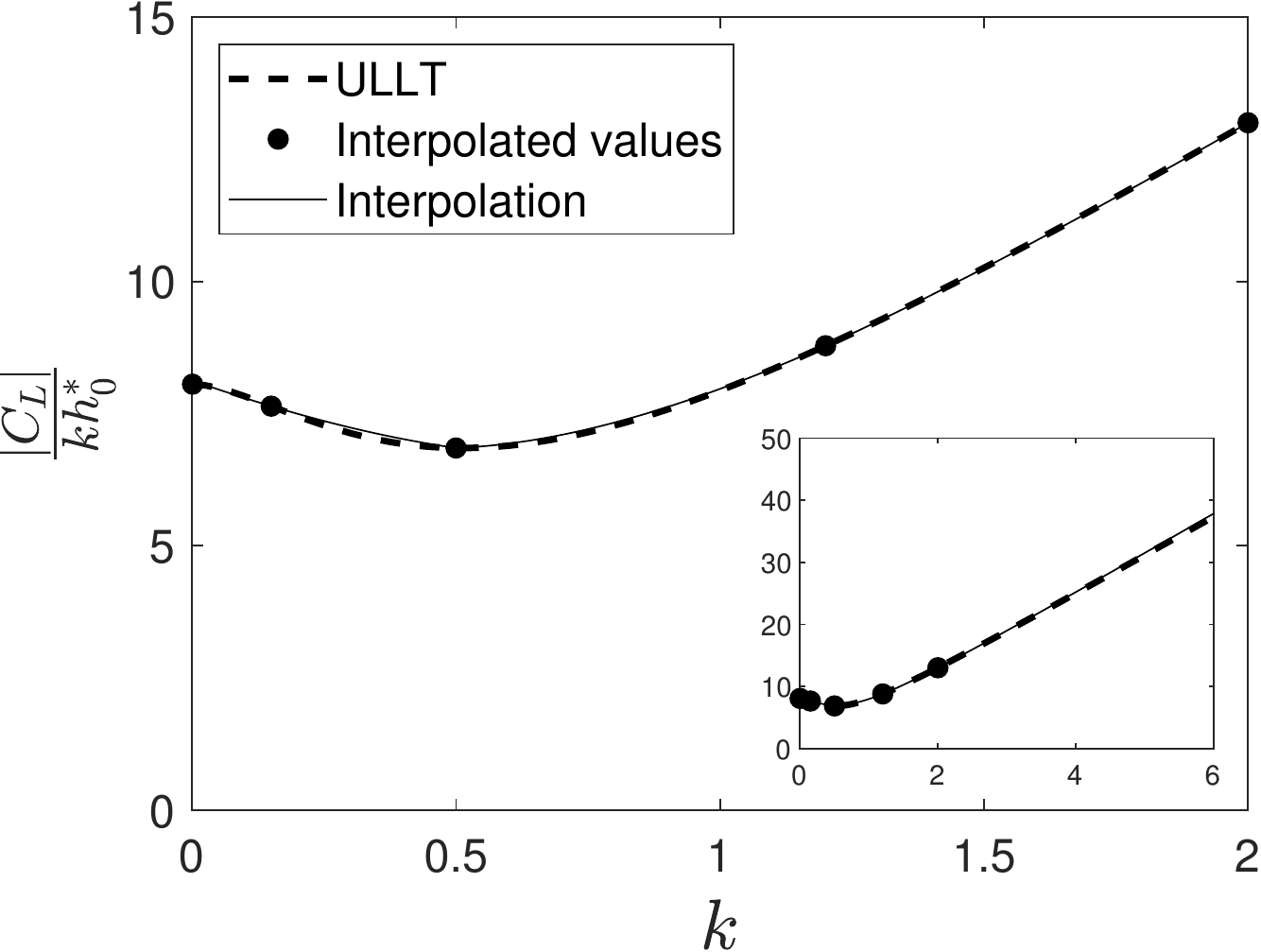}}
    \subfigure[Pitch $C_L$]{
	\label{fig:pitch_CL_interp}
    	\includegraphics[width=0.4\textwidth]{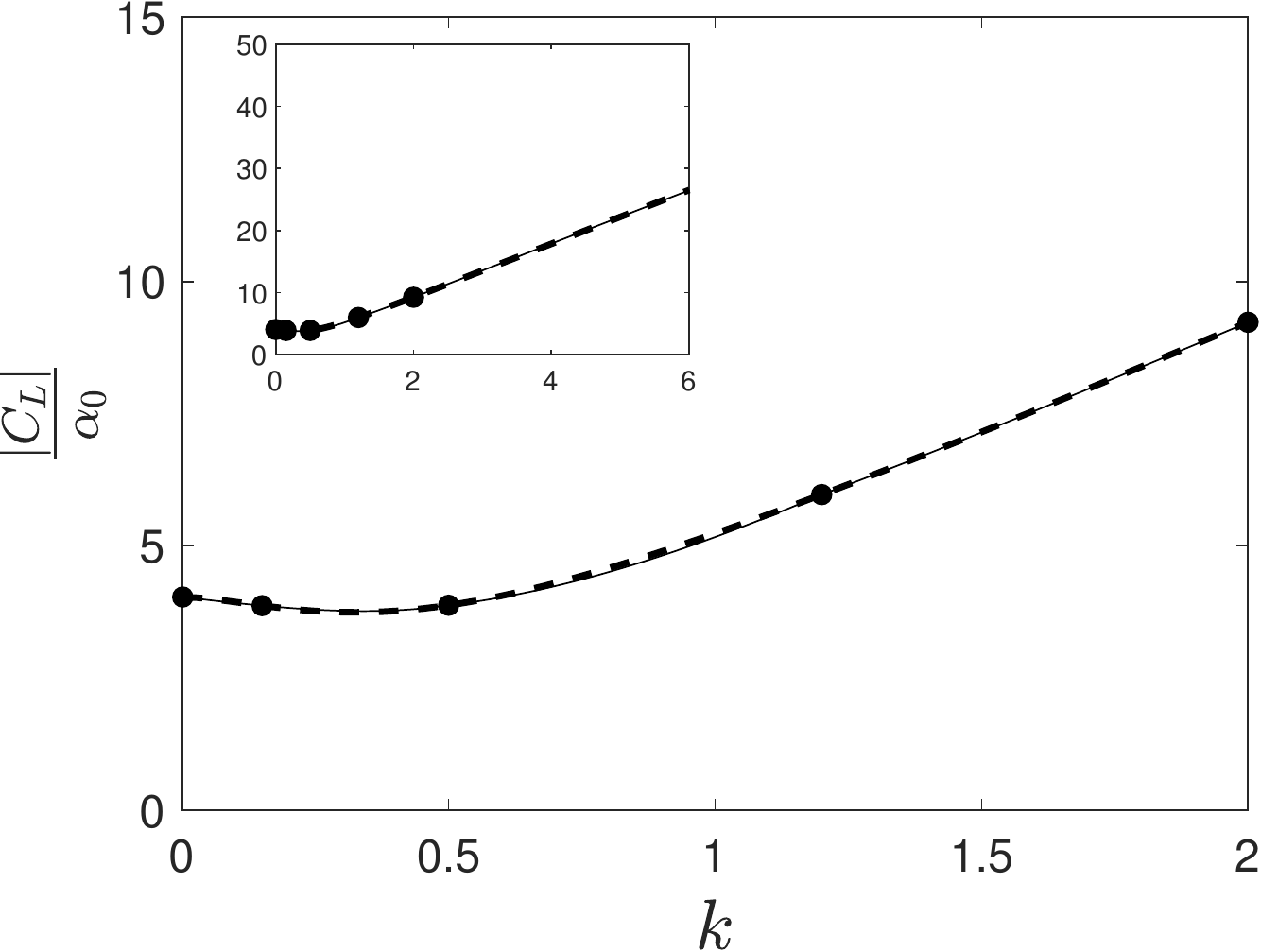}}

    \subfigure[Heave $C_M$]{
	\label{fig:heave_CM_interp}
    	\includegraphics[width=0.4\textwidth]{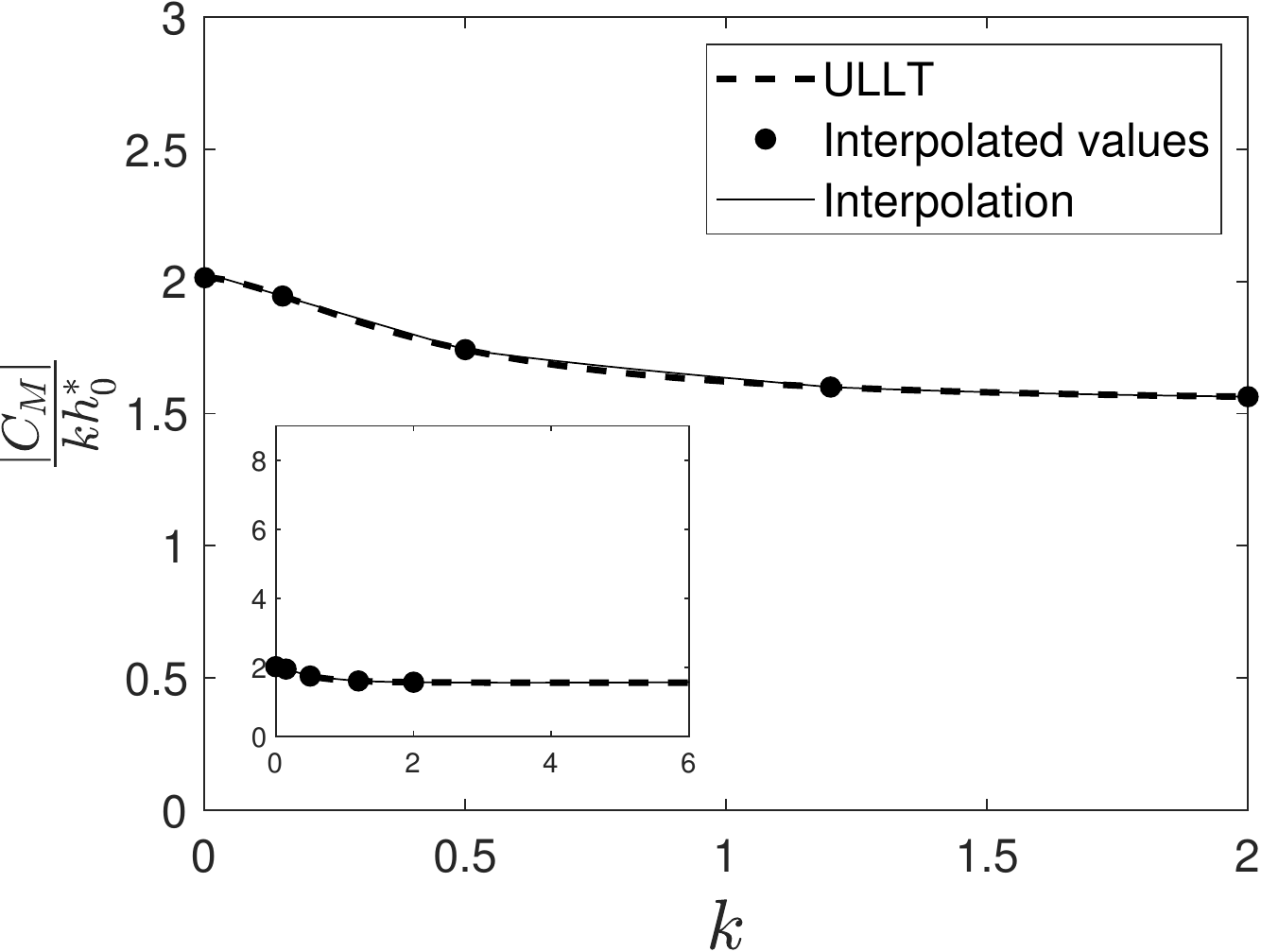}}
    \subfigure[Pitch $C_M$]{
	\label{fig:pitch_CM_interp}
    	\includegraphics[width=0.4\textwidth]{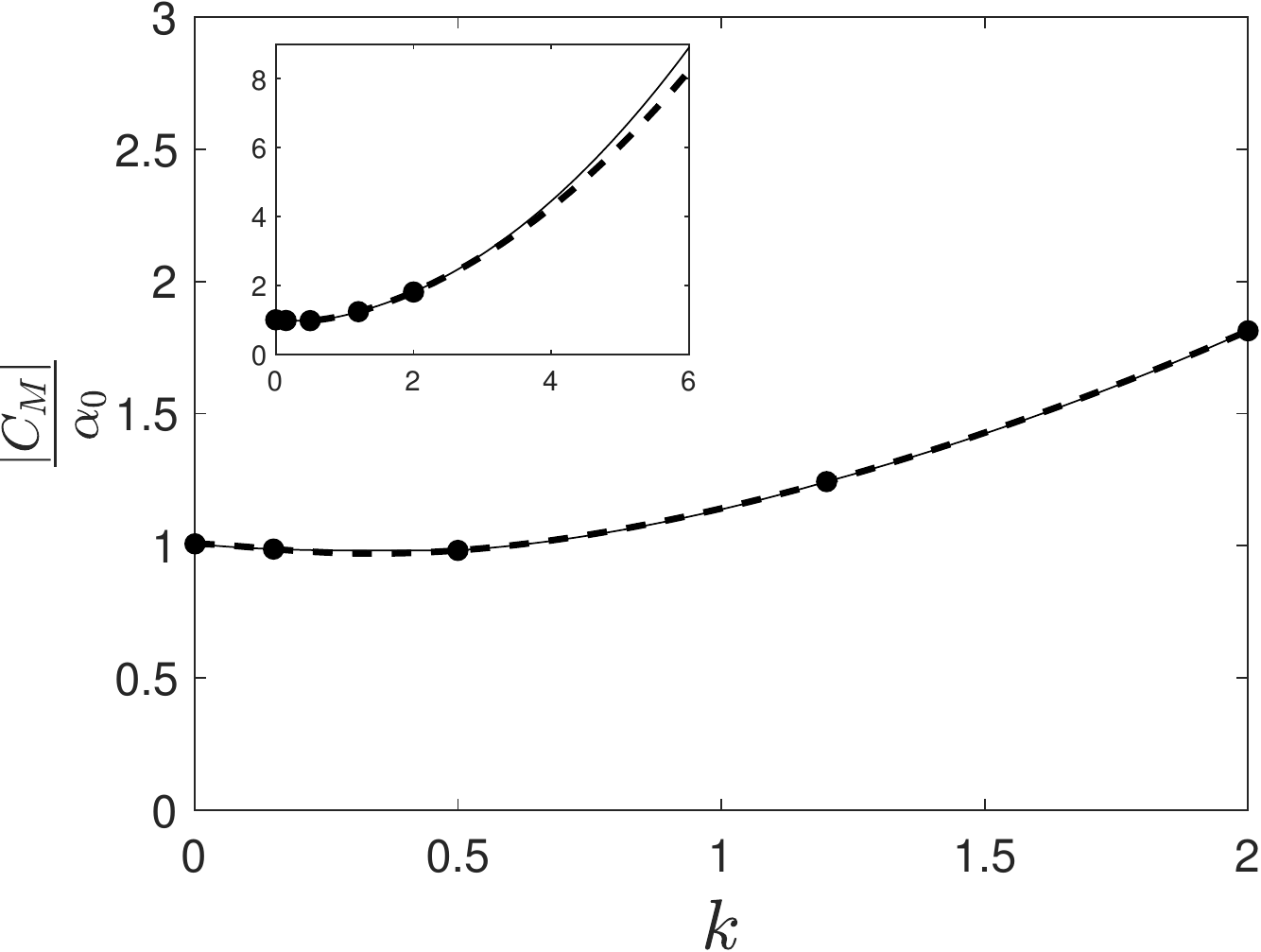}}

	\caption{Interpolation of lift and moment coefficients with respect to reduced frequency for
	a rectangular aspect ratio 4 wing. Moments are taken about the mid-chord.
	Inset shows extended frequency range result.}
	\label{fig:frequency_domain_interpolation}	
\end{figure}

This interpolation strategy is surprisingly effective. 
Figure~\ref{fig:frequency_domain_interpolation} shows 
a 5 frequency interpolation for an aspect ratio 4 rectangular wing 
with the ULLT evaluated at 
$k = \{0.001, 0.15, 0.5, 1.2, 2.0\}$. This set of reduced frequencies 
is used for constructing the interpolation of all cases in this paper.

Interpolation provides excellent results even if only a small number of frequencies are evaluated
for a given wing planform, even though extrapolation is required for high frequencies.
For any given wing, this frequency domain result need only be obtained once.

Traditionally, time-domain schemes have combined step-function responses (indicial functions) with
Duhamel integrals to obtain a response with respect to arbitrary inputs. 
This is a convolution of the input kinematics with the step response. To avoid the
challenge of obtaining the step response, in the UCoFD method the convolution is performed
in the frequency domain.
This is done by taking the Fourier transform of the input kinematics, taking the convolution of
this with the force response, and then taking the inverse Fourier transform. For a heave displacement
 $h(t)$ or pitch angle $\alpha(t)$ the following can be applied:

\begin{align}
	C_{L_h}(t) = \mathcal{F}^{-1}\left( \mathcal{F}(h(t))  \cdot C_{L_h}(\omega) \right) ,\\
	C_{M_h}(t) = \mathcal{F}^{-1}\left( \mathcal{F}(h(t))  \cdot  C_{M_h}(\omega) \right) ,\\
	C_{L_\alpha}(t) = \mathcal{F}^{-1}\left( \mathcal{F}(\alpha(t))   \cdot C_{L_\alpha}(\omega) \right) ,\\
	C_{M_\alpha}(t) = \mathcal{F}^{-1}\left( \mathcal{F}(\alpha(t))  \cdot  C_{M_\alpha}(\omega) \right) . 
\end{align}

This can be done quickly and easily using readily available fast Fourier transform software packages. Note
that for negative angular frequencies, the complex conjugate of the coefficient can be used as 
$C_{L/M}(-\omega) = \overline{C_{L/M}(\omega)}$.

By taking the Fourier transform of the input kinematics we assume that they
are periodic. In practice if the time window passed into the Fourier transform
is sufficiently large, this does not matter. The only caveat is that the displacement
of the wing at the beginning and end of the time window should be the same.

\section{Results and discussion}
\label{sec:results}

Currently, strip-theory is often used to obtain solutions to time-domain unsteady finite wing problems quickly.
In this section frequency-domain unsteady lifting-line theory is applied to time-domain
problems using the UCoFD method, and the results obtained are compared against those from computational
fluid dynamics and the often-used strip theory for both inviscid and low-Reynolds-number cases (Re$=$\num{10000}). 
The CFD setup is detailed in Sec.~\ref{sec:cfd},
with the Euler setup specific aspects given in Sec.~\ref{sec:euler_cfd}, and 
the low Reynolds number specific aspects given in Sec.~\ref{sec:low_re_cfd}.
The CFD results obtained from these setups serve as a reference against
which the results obtained from the UCoFD method and strip-theory can be compared.
The six cases studied are shown in Table~\ref{tab:cases}.

These cases are based on the canonical ramp-hold-return kinematics of
Ol et al.~\cite{Ol2010}. These kinematics were chosen to demonstrate time-domain problems 
applicable to flapping flight where both circulatory and added-mass effects are prevalent.
Case 1 (Sec.~\ref{sec:pitch}) is a pitch ramp-hold-return in the Euler regime (Re$=\infty$), for which the UCoFD
method is expected to work well. Next Case 2, in Sec.~\ref{sec:heave}, is a Euler regime heave velocity ramp-hold-return
that introduces complications with respect to having a non-zero final displacement and being
less smooth. Finally, in Sec.~\ref{sec:low_re}, Cases 3a-3d are pitch ramp-hold-returns at Re$=$\num{10000} 
(a regime representative of modern engineering problems).
Pitch ramp-hold-return kinematics are applied to wings of aspect ratio 6 and 3, for both small-amplitude
$\alpha_{\text{max}}= \ang{3}$ and large-amplitude $\alpha_{\text{max}}= \ang{25}$ kinematics.

\begin{table}
	\centering
	\caption{Cases for comparison between UCoFD, CFD and strip theory.}
	\begin{tabular}{l l l l}
	\toprule
	Case & Reynolds No. & Aspect ratio & Kinematics \\
	\midrule
	1 & $\infty$ & 4 & Small-amplitude smooth pitch ramp-hold-return\\
	2 &$\infty$ & 4 & Small-amplitude non-smooth heave velocity ramp-hold-return\\
	3a& \num{10000} & 6 & Small-amplitude smooth pitch ramp-hold-return\\
	3b& \num{10000} & 6 & Large-amplitude smooth pitch ramp-hold-return\\
	3c& \num{10000} & 3 & Small-amplitude smooth pitch ramp-hold-return\\
	3d& \num{10000} & 3 & Large-amplitude smooth pitch ramp-hold-return\\
	\bottomrule
	\end{tabular}
	\label{tab:cases}
\end{table}

\subsection{Computational fluid dynamics}
\label{sec:cfd}

The open-source CFD toolbox OpenFOAM was used to 
perform numerical computations. For both Euler and low-Reynolds-number case,
a body-fitted, structured computational mesh is moved according to
prescribed heave and pitch kinematics, and the time-dependent
governing equations are solved using a finite volume method. A
second-order backward implicit scheme is used to discretize the time
derivatives, and second-order limited Gaussian integration schemes are
used for the gradient, divergence and Laplacian terms. 
Pressure-velocity coupling is achieved using
the pressure implicit with splitting of operators (PISO) algorithm.

\subsubsection{Euler cases}
\label{sec:euler_cfd}
The Euler cases are modeled using an in-house setup that has previously been
used to validate unsteady potential flow solutions for an airfoil~\cite{ramesh2020leading}
and for rectangular wings~\cite{Bird2021b}.
This article shares the detailed setup and meshes of the latter.
The setup is intended to model the regime for which the current ULLT
was derived - that is, inviscid, incompressible flow with small
perturbations. Laminar flow is considered with kinematic viscosity set to
zero, and a slip boundary condition is employed for the moving wing
surface. 

An aspect ratio 4 rectangular wing with squared off tips is considered. 
A NACA0004 section is chosen to best match the theoretical
assumptions of thin section and Kutta condition at the trailing edge.
Since the kinematics are
symmetric about the wing center, the cylindrical O-mesh
is constructed for only half the wing. The
mesh has $160$ cells around the wing section, the
resolution increasing near the leading and trailing edges. 
In the wall-normal
direction, the mesh has $115$ cells with the far-field extending $20$ chord
lengths in all directions around the section. In the spanwise
direction, the mesh has $199$ cells over the semispan of the wing with the resolution
increasing near the wingtip. The spanwise domain extends $100$ cells beyond the wingtip,
to a total of 5 chord lengths.

\subsubsection{Low Reynolds number cases}
\label{sec:low_re_cfd}

The Re$=$\num{10000} cases were computed following the experimentally-validated 
RANS method used in Bird et al.~\cite{Bird2021c}, albeit with
different kinematics.
The Spalart-Allmaras (SA) turbulence
model~\cite{Spalart1992} is used for turbulence closure, with
the trip terms in the original SA model turned off.
For the low Reynolds number cases considered in this
research, the effects of the turbulence model are confined to the shed
vortical structures and wake. 

The meshes are the same as those used in Bird et al.~\cite{Bird2021c}.
In this paper, two rectangular wings of chord length $c= 0.1$ m and
aspect ratios 6 and 3 are considered.
An O-mesh topology was used with 116 cells in the chordwise direction.
The mesh was finer near the leading and trailing edges. 
Since the pitch kinematics are symmetrical about the wing center, only
half the wing was meshed.
The aspect ratio 6 and 3 half-wing meshes had 211 and 105 cells respectively in the spanwise
direction. The spanwise
domain extends $4$ chord lengths beyond the wingtip with an average
spacing of $21$ cells per chord length in this region. In the wall-normal direction,
cell spacing begins at $1.5\times10^{-5}$ m next to the wall ($y+ < 1$) and extends a distance 
of $11.5$ chord lengths away from the wing with an average density of 16.3 cells per
chord length.
The simulations were
carried out at a free stream velocity $U_\infty=0.1$ m$/$s and kinematic
viscosity $10^{-6}$ m$^2/$s to obtain a chord-based Reynolds
number of \num{10000}.

\subsection{Case 1: a returning pitch ramp in the Euler regime}
\label{sec:pitch}

\begin{figure}
	\centering
    	\includegraphics[width=0.4\textwidth]{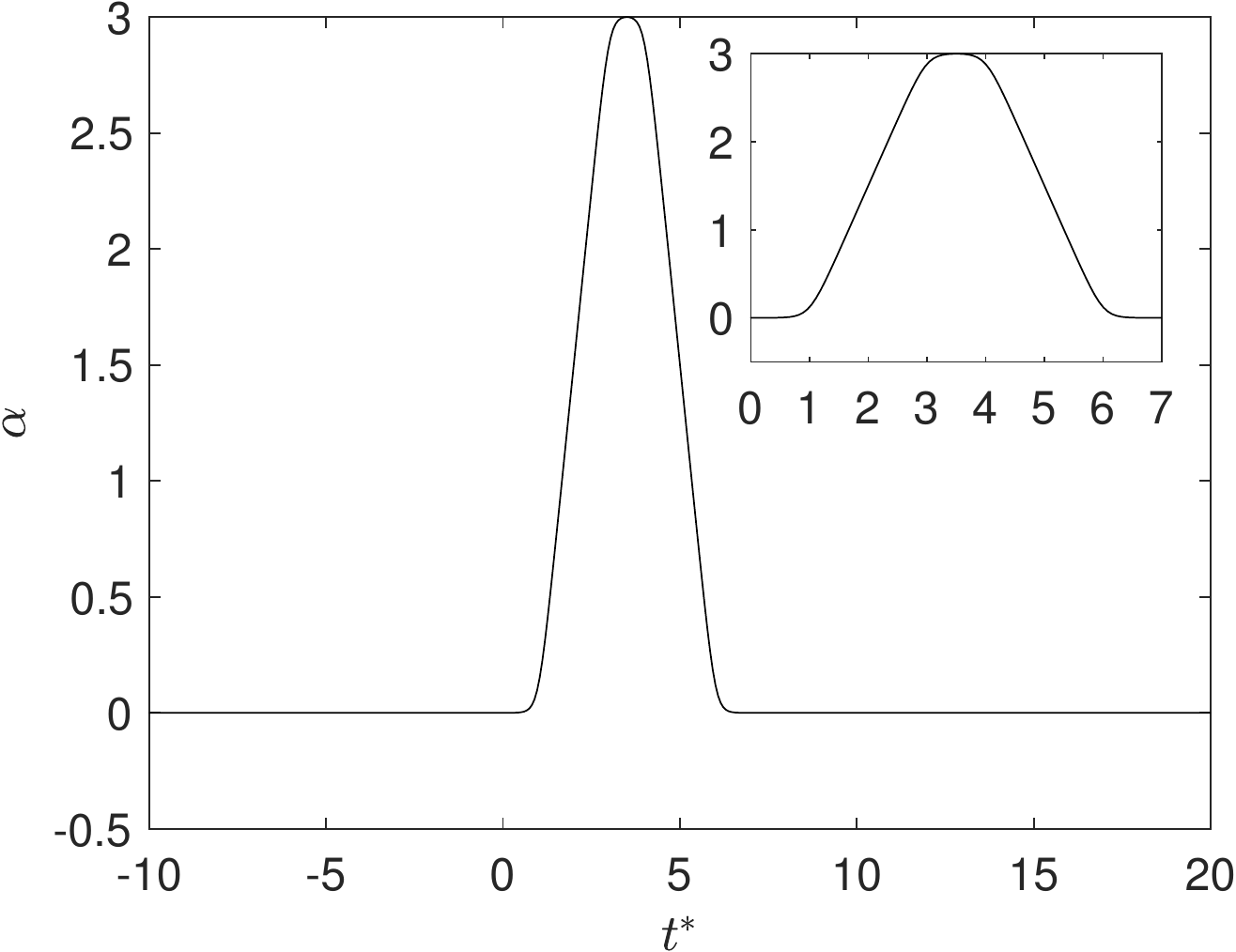}
	\caption{Leading edge pitch ramp kinematics.}
	\label{fig:pitch_displacement}
\end{figure}

A canonical pitch ramp-hold-return motion \cite{Ol2010} is expressed as
\[
	\alpha(t) = \frac{P}{a\bar{c}}\left[ \frac{\cosh(aU_\infty(t-t_1)/\bar{c}) \cosh(aU_\infty(t-t_4)/\bar{c})}{\cosh(aU_\infty(t-t_2)/\bar{c})\cosh(aU_\infty(t-t_3)/\bar{c})} \right]
\]
where $a=\pi^2/(4(t_2-t_1)(1-\sigma))$. This ramp-hold-return motion has the advantage that
the start and end displacement is identical - unlike for the heave velocity ramp-hold-return
studied in the next subsection.
Here $P$ set to give $\alpha_{\text{max}}= \ang{3}$, 
and the time parameters are set to $t_1^*=1$, $t_2^*=3$, $t_3^*=4$ and $t_4^*=6$,
where $t^* = t U_\infty / \overline{c}$.
The smoothness of the curve is dictated by $\sigma=0.5$. The 
kinematics are shown in Fig.~\ref{fig:pitch_displacement}.
A window of $t^*=[-10,20]$
is used here, despite the input signal of interest being in $t^*=[0,7]$ (inset). 
A large window is required in comparison to that
of the input signal since the Fourier transform with limited lower bounds on input frequency
effectively assumes that the input is periodic. Consequently, a large window avoids the
issues introduced by periodicity.

These input kinematics were evaluated at 2048 evenly spaced points
in the $t^*=[-10,20]$ window. For a problem involving a rectangular aspect ratio 4 wing of
 chord $c=$ \SI{0.0762}{\metre} in a freestream of  $U_\infty=$ \SI{0.1312}{\metre\per\second},
this gives a sample interval of $\Delta t \approx$ \SI{0.015}{\second}, resulting in
a Nyquist critical frequency of $\omega_c = 2\pi/2\Delta t = $ \SI{369}{\radian \per \second} which corresponds to a critical chord 
reduced frequency of $k_c = 107$. This is sufficiently high to capture
the highest frequency present in the input signal.

\begin{figure}
\centering
    \subfigure[$C_L$]{
	\label{fig:pitch_ramp_CL}
    	\includegraphics[width=0.4\textwidth]{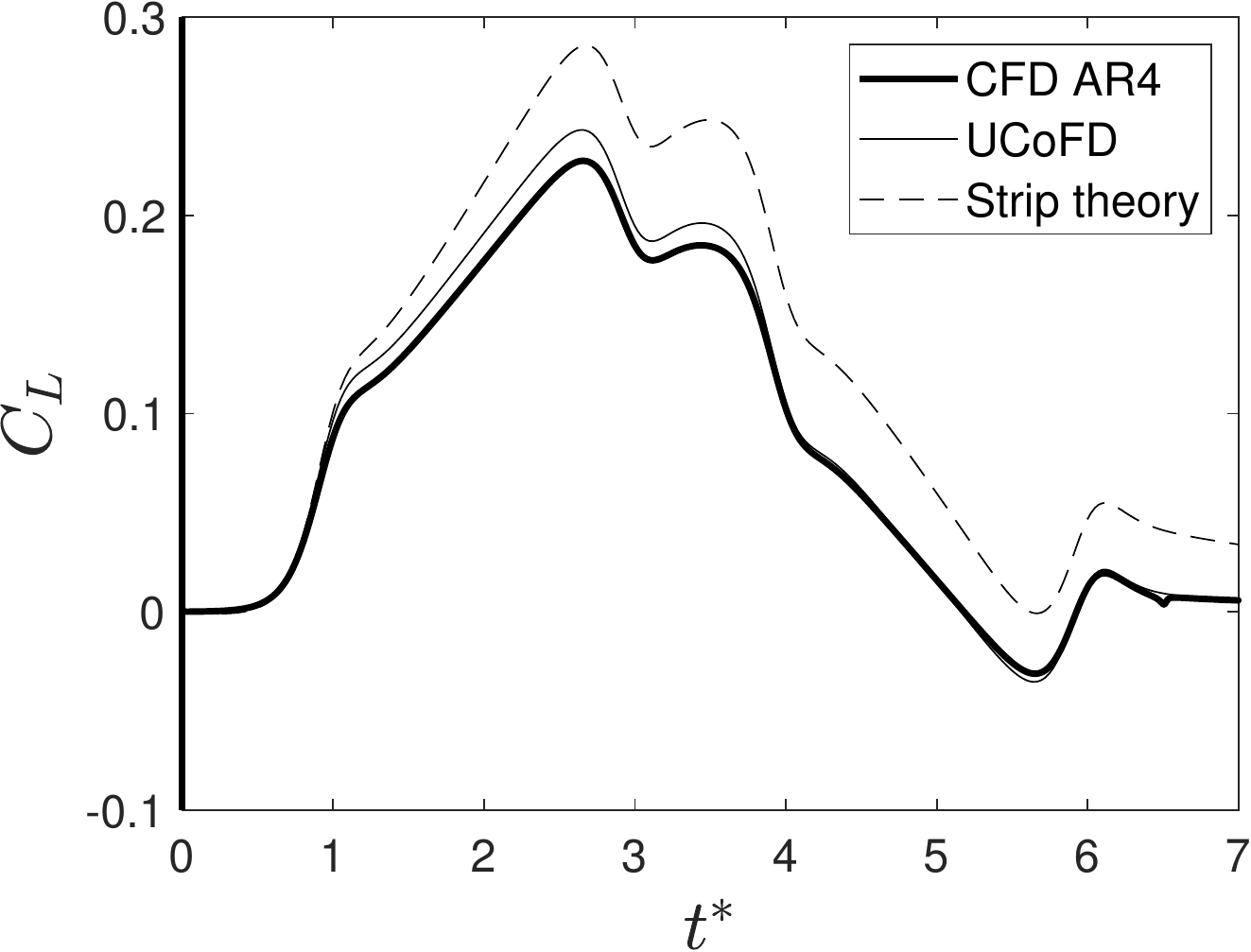}}
    \subfigure[Leading-edge $C_M$]{
	\label{fig:pitch_ramp_CM}
    	\includegraphics[width=0.4\textwidth]{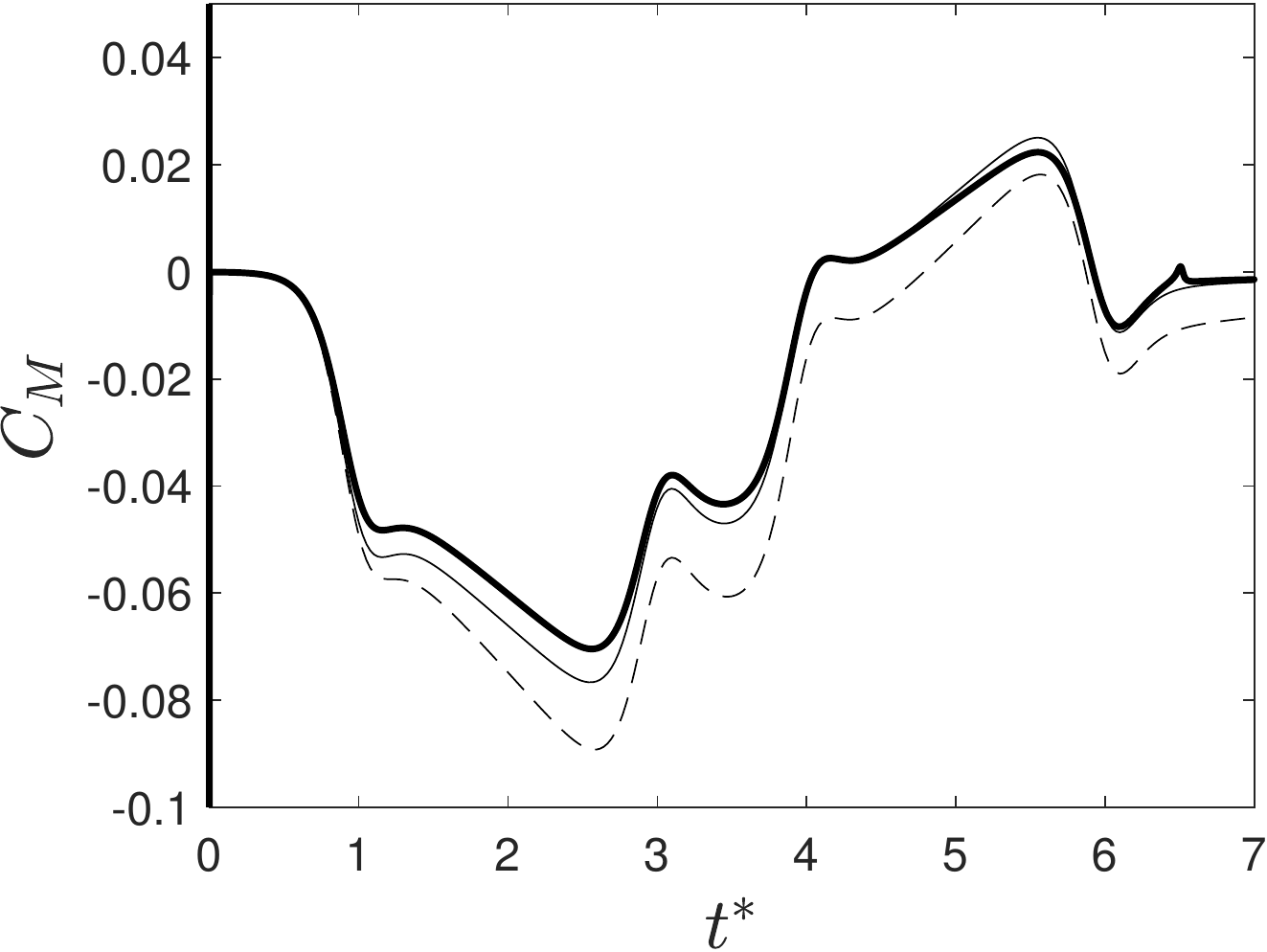}}

	\caption{Comparison of UCoFD, CFD and strip theory results for an aspect ratio 4 rectangular wing undergoing
	a smooth pitch ramp maneuver. }
	\label{fig:eldredge_pitch_CL_CM}	
\end{figure}

The $C_L$ and $C_M$ results obtained using CFD, UCoFD and strip theory
are shown in Fig.~\ref{fig:eldredge_pitch_CL_CM}.

The CFD results in Fig.~\ref{fig:pitch_ramp_CL} show the lift increasing and then returning
to zero, roughly in  line with the input ramp kinematics. Two mechanisms create
lift. Firstly, the acceleration of the wing leading to added mass effects. The
positive acceleration at $t^*=1$ leads to a very rapid initial increase in lift. The
negative acceleration either side of the peak at $t^*=3$ and $t^*=4$ leads to
a reduction in lift. The positive acceleration at $t^*=6$, from negative to zero rate of pitch,
leads to an increase in lift. Circulatory effects lead to the overall shape of the curve, where
lift is approximately proportional to the effective angle of attack~\cite{Ramesh2020}. The circulatory and acceleratory
effects sum to the more complex final lift curve.

The shape of the $C_L$ curve is matched excellently
by the unsteady lifting-line theory result, although the lift peak is slightly
overestimated. In contrast, strip 
theory significantly overestimates the peak lift and introduces
a residual lift that does not reflect the results obtained from the CFD.

The moment coefficients, shown in Fig.~\ref{fig:pitch_ramp_CM}, tell a similar
story. The CFD result shows the $C_M$ to be a combination of acceleratory 
effects, which lead to rapid changes in $C_M$, and circulatory effects, which
lead to the overall shape of the curve.
This is reflected almost perfectly by the ULLT result, except for a small overestimation
of the peak value of $C_M$. Again, strip theory significantly overestimates the peak
value of the coefficient, and fails to return to zero as quickly as the CFD result.

For this pitch case the assumptions made in the derivation of the ULLT 
were satisfied (excepting the rectangular wing shape), and the fast Fourier transform 
could be applied without complication. This leads to excellent results from the
UCoFD method.
Next, a more challenging case will be examined, introducing two 
complications. Firstly, a less smooth input function will be used. This introduces
higher frequency terms that may introduce errors to the result of the ULLT. And
secondly, the input kinematics will not return to zero, and an artificial return 
ramp must be introduced. 

\subsection{Case 2: a non-returning heave velocity ramp in the Euler regime}
\label{sec:heave}

In this subsection, the canonical ramp-hold-return motion is once
again used \cite{Ol2010}, but instead of being applied to pitch angle
it is applied to heave velocity. As a consequence, the net heave displacement is
non-zero. 

The ramp-hold-return motion is defined by the heave velocity $\dot h = dh/dt$ as
\[
	\dot h= \frac{P}{a}\left[ \frac{\cosh(aU_\infty(t-t_1)/\bar{c}) \cosh(aU_\infty(t-t_4)/\bar{c})}{\cosh(aU_\infty(t-t_2)/\bar{c})\cosh(aU_\infty(t-t_3)/\bar{c})} \right],
\]
where $a=\pi^2/(4(t_2-t_1)(1-\sigma))$ and $P$ set to give $\dot h_\text{max}= -0.05\bar{c}$, 
and the timing parameters are $t_1^*=1$, $t_2^*=3$, $t_3^*=4$ and $t_4^*=6$.
The smoothness of the curve is set as $\sigma=0.888$, which results in less
smooth motion in comparison to the pitch ramp used for Case 1 in Sec.~\ref{sec:pitch}. 

The displacement at the end of the heave velocity ramp is not equal to that
at the beginning. Consequently, for the UCoFD method using a window $t^*=[-10,35]$, there is a
step from $t^*=35$ to $t^*=-10$. This must be mitigated to obtain good results.

To do remove the jump, a smooth return function is inserted. Here
a quadratic return ramp is used. This ramp is given by
\begin{equation}
	g(t) = 
	\begin{cases}
	    1,& \text{if } t \leq t_{s_0}\\
	    1-2\left(\frac{t-t_{s_0}}{t_{s_1}-t_{s_0}}\right)^2,& \text{if } t_{s_0}  < t \leq (t_{s_0}+t_{s_1})/2\\
	    2\left(\frac{t_{s_1}-t}{t_{s_1}-t_{s_0}}\right)^2,& \text{if } (t_{s_0}+t_{s_1})/2 < t \leq t_{s_1}\\
	    0,& \text{if } t > t_{s_1}
	\end{cases}
\end{equation}

To obtain a displacement function which returns to zero, $h'(t)$, the original
heave ramp is multiplied by the return function, giving $h'(t) = h(t) g(t)$.
For a problem where the heave ramp is in $t^*=[0,7]$ and the investigation window
is in $t^*=[-10,35]$, the return ramp used parameters $t_{s_0}^*=10$ and $t_{s_1}^*=20$.
This places it immediately after the results of interest, allowing a large separation between
the return ramp and, given the results are periodic, the beginning of the input kinematics.
Again, these kinematics are applied to a rectangular wing of aspect ratio 4, with chord $c=$ \SI{0.0762}{\metre}
and a NACA0004 section in a freestream of $U_\infty=$ \SI{0.1312}{\metre\per\second},
with results obtained using a 2048-sample fast Fourier transform.

\begin{figure}
\centering
    \subfigure[Heave kinematics]{
	\label{fig:heave_kinem}
    	\includegraphics[height=0.35\textwidth]{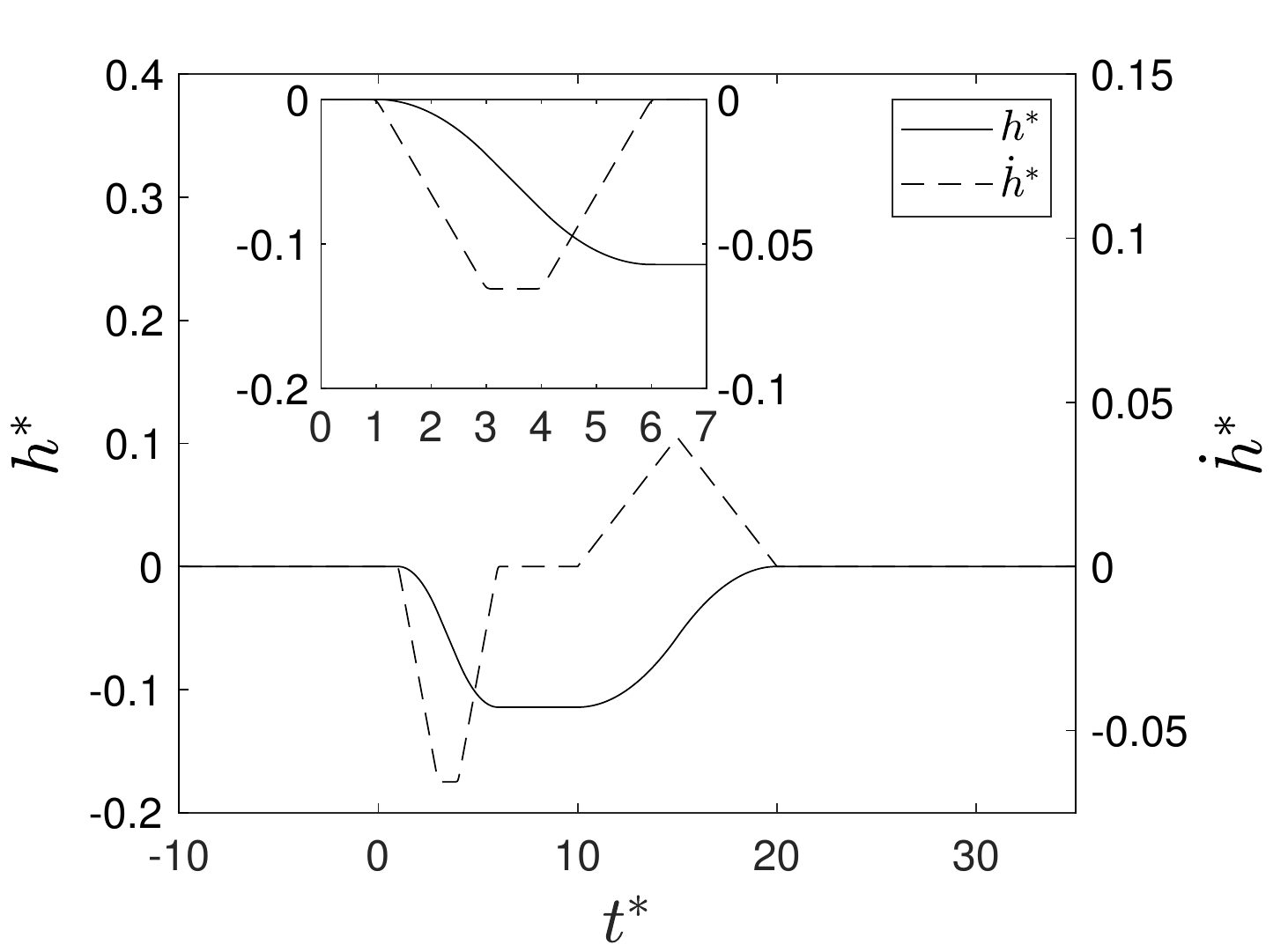}}
    \subfigure[Force coefficients]{
	\label{fig:heave_CL_CM_zoom}
    	\includegraphics[height=0.35\textwidth]{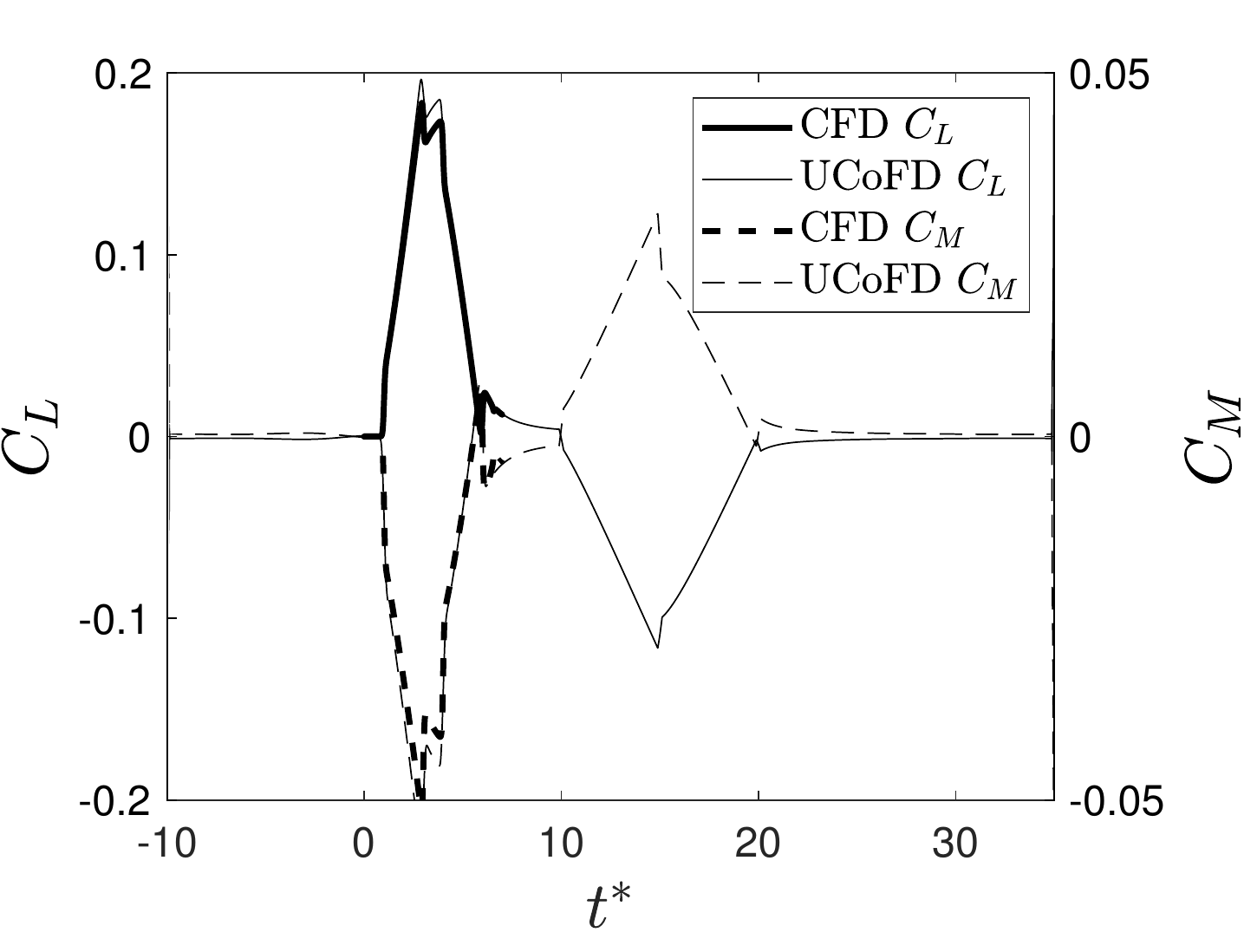}}

	\caption{Heave ramp kinematics and results for an aspect ratio 4 rectangular wing.}
	\label{fig:eldredge_heave_zoom}	
\end{figure}

The input kinematics and the results obtained for the rectangular wing are show in
Fig.~\ref{fig:eldredge_heave_zoom}

In Fig.~\ref{fig:heave_kinem}, the heave ramp of interest can be seen in $t^*=[0,7]$ (inset). 
This ramp results in a non-zero displacement by $t^*=7$.
As discussed earlier, a return ramp is introduced to avoid a step between $t^*=35$ and 
$t^*=-10$. This ramp is separated from the kinematics of interest. 
The reason for this can be seen at $t^*=20$ in
Fig.~\ref{fig:heave_CL_CM_zoom}, where
it takes time for the forces on the wing to return to zero after the return ramp has finished. 

\begin{figure}
\centering
    \subfigure[$C_L$]{
	\label{fig:eldredge_heave_CL}
    	\includegraphics[width=0.4\textwidth]{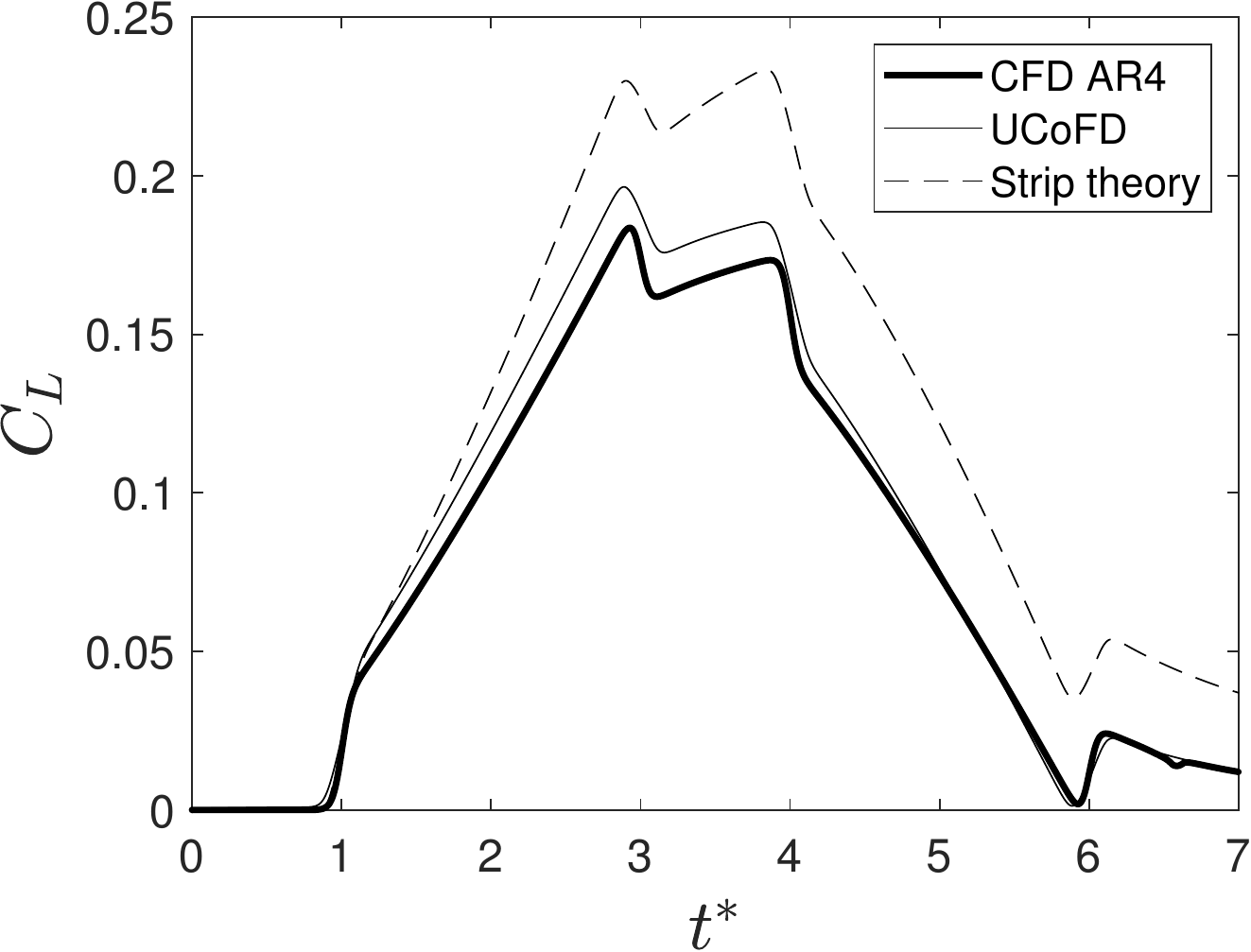}}
    \subfigure[$C_M$]{
	\label{fig:eldredge_heave_CM}
    	\includegraphics[width=0.4\textwidth]{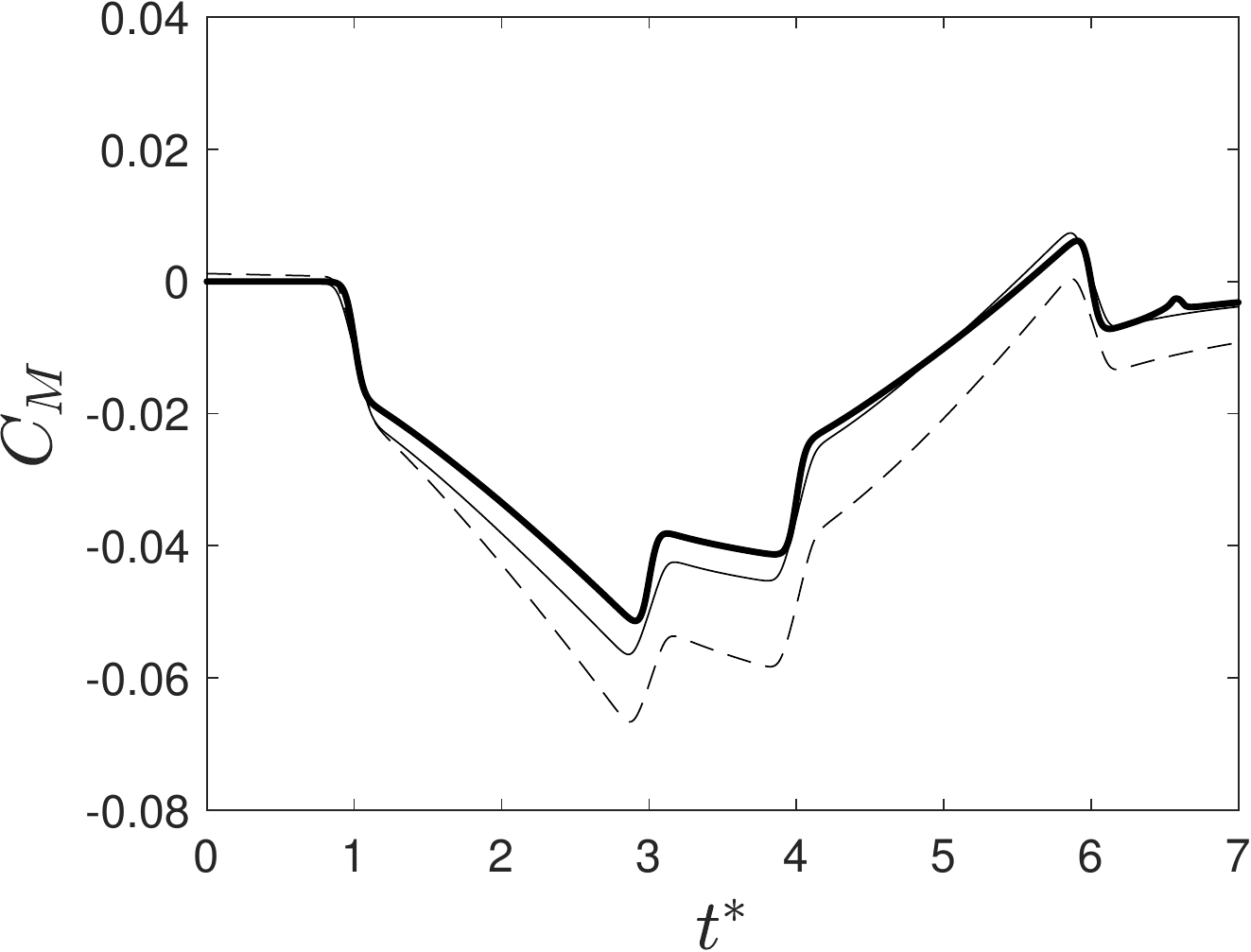}}

	\caption{Comparison of UCoFD, CFD and strip theory results for an aspect ratio 4 rectangular wing undergoing
	a non-smooth heave velocity ramp maneuver.}
	\label{fig:eldredge_heave_CM_CL}	
\end{figure}

Detailed force coefficient results in the interval of interest are shown in Fig.~\ref{fig:eldredge_heave_CM_CL}.
The CFD results for both the $C_L$ and $C_M$ curves shows an increase in
force, a plateau, and a reduction in force. The non-smoothness of the input kinematics
is reflected by the sudden change in $C_L$ and $C_M$ found at $t^*=\{1,3,4,7\}$, where the
rapid change in the acceleration of the wing leads to changes in the forces due to added-mass effects.

The ULLT / convolution in frequency domain method again predicts the CFD result with good accuracy. The overall shape
of the $C_L$ and $C_M$ curves are well predicted, including the strength of the 
peaks resulting from added mass effects. This suggests that the theoretical asymptotic
limitation of the underlying frequency-domain ULLT is of little practical consequence.
Again, the ULLT slightly over-predicts the peak values of the forces. However,
this error is small, especially when compared to the error introduced if strip-theory were
to be used instead.

It has been demonstrated how a frequency-domain ULLT can be applied successfully
to time-domain problems using the UCoFD method,
and that it gives superior results to strip theory. 
However, the analytical frequency-domain ULLTs to which this technique
can be applied almost universally assume potential flow and small amplitude kinematics.
In the next section, the results obtained when these assumptions are broken are examined.

\subsection{Cases 3a-3d: large amplitude pitch ramp at Re=\num{10000}}
\label{sec:low_re}

To solve for the aerodynamics in research areas such as micro air vehicles, 
unmanned aerial vehicles~\cite{Williams2002} or energy harvesting devices~\cite{Rostami2017},
solutions for low-Reynolds number, high-amplitude
kinematics problems are required. 
The linear, potential-flow based model
used in the derivation of Sclavounos' ULLT obtains surprisingly good results
for frequency-domain problems in this regime \cite{Bird2021c} given that the assumptions
used in its derivation are broken. Here, time-domain cases are explored using the UCoFD method.

Two rectangular wings, one of aspect ratio 6 (Cases 3a and 3b) and the other of aspect ratio 3 (Cases 3c and 3d),
undergo a pitch ramp-hold-return maneuver at a Reynolds number of 
\num{10000}. This Reynolds number was chosen since it is representative
of the low-Reynolds-number regime of the above applications.
The wing has chord $c=$ \SI{0.1}{\meter}, a NACA0008 airfoil section, 
and the free stream velocity is $U_\infty=$ \SI{0.1}{\meter\per\second}.

The pitch ramp-hold-return maneuver is similar to that of Case 1 in Sec.~\ref{sec:pitch}. The wing pitches about
its leading edge with $P$ set to give both $\alpha_{\text{max}}= \ang{3}$ (Cases 3a and 3c) and $\alpha_{\text{max}}= \ang{25}$ (Cases 3b and 3d), 
and the timing parameters are set to $t_1=1$, $t_2=3$, $t_3=4$ and $t_4=6$.
The smoothness of the curve is given by $\sigma=0.5$. 

\begin{figure}
\centering
    \subfigure[Aspect ratio 6 $C_L$]{
	\label{fig:eldredge_pitch25_ar6_CL}
    	\includegraphics[width=0.4\textwidth]{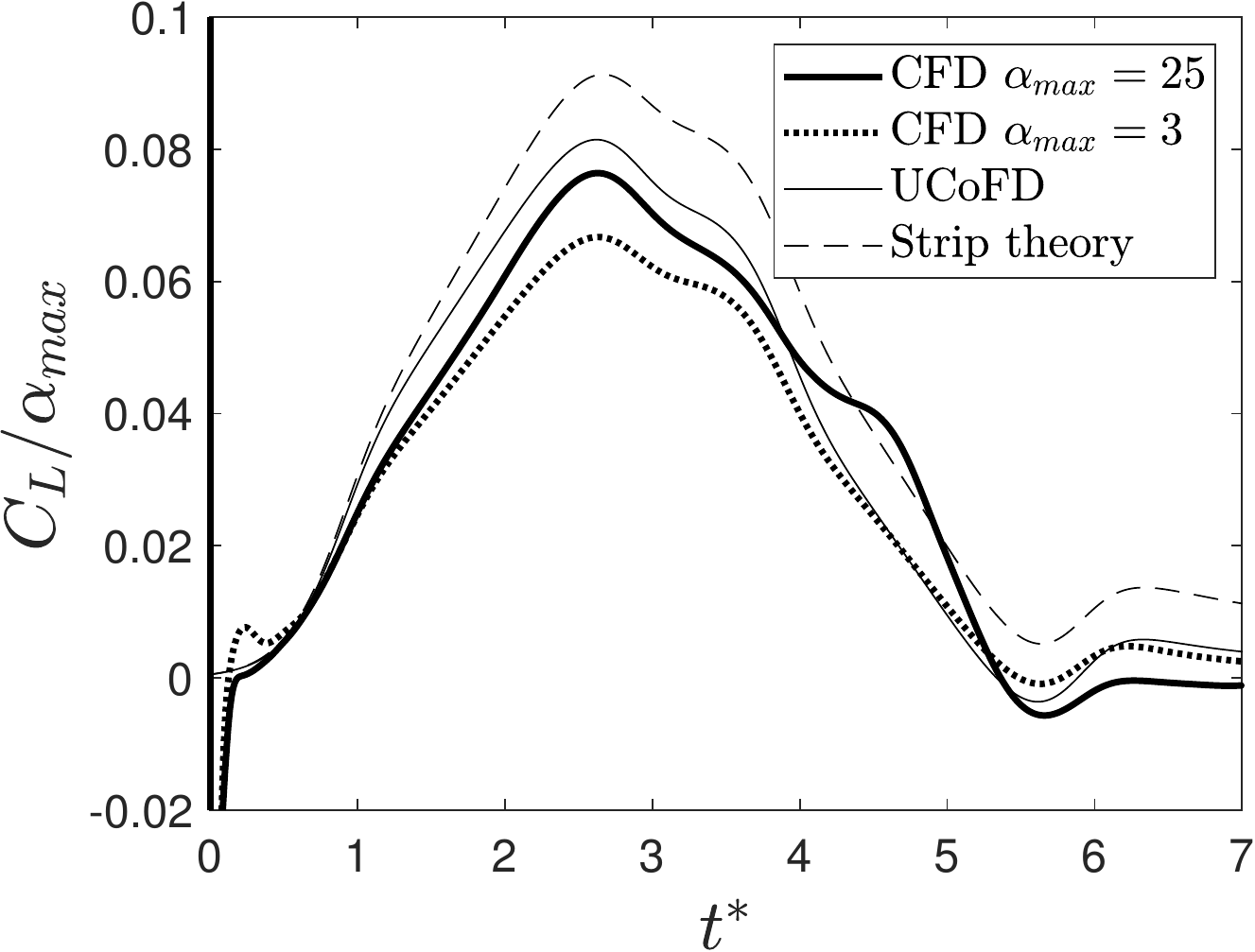}}
    \subfigure[Aspect ratio 6 $C_M$]{
	\label{fig:eldredge_pitch25_ar6_CM}
    	\includegraphics[width=0.4\textwidth]{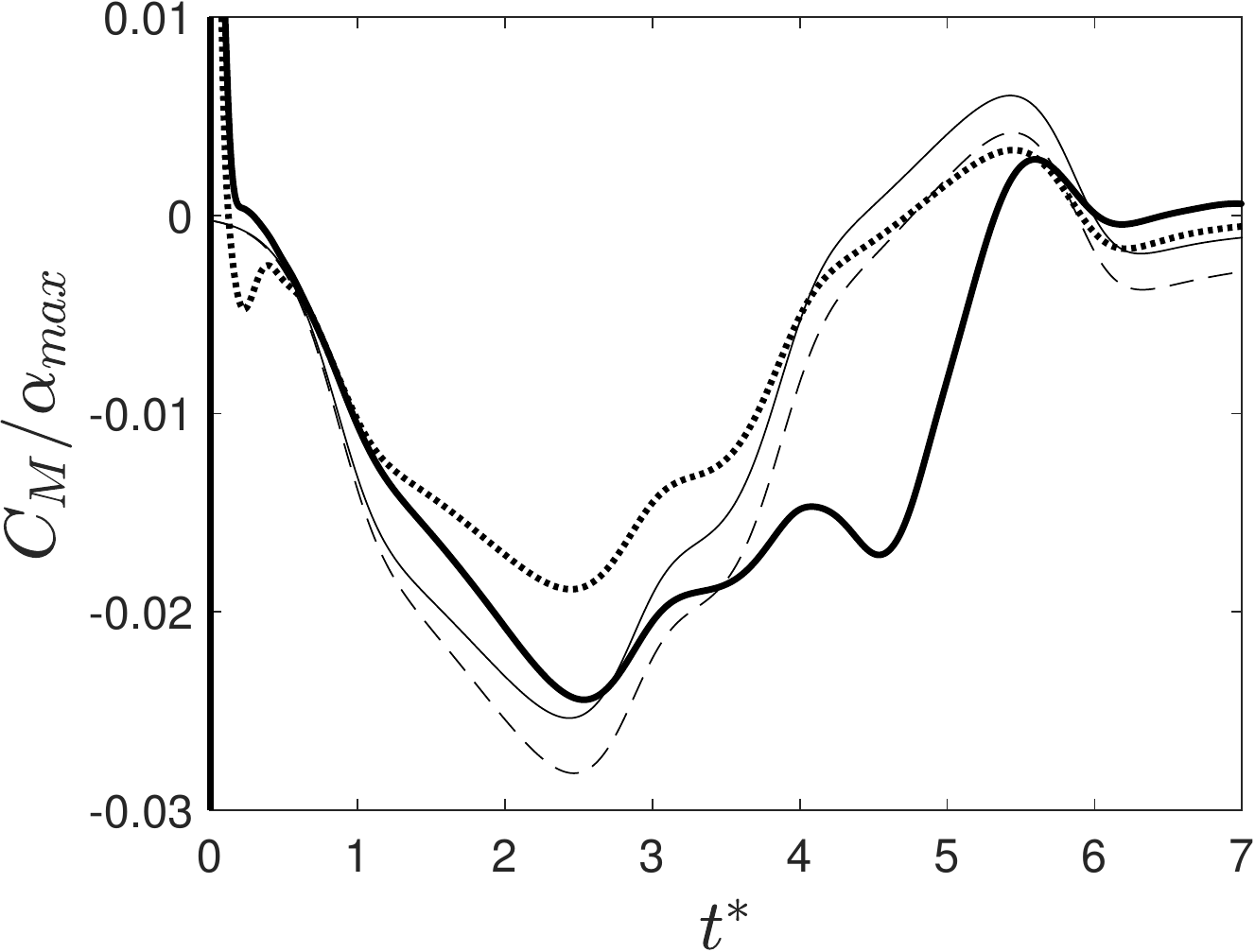}}

    \subfigure[Aspect ratio 3 $C_L$]{
	\label{fig:eldredge_pitch25_ar3_CL}
    	\includegraphics[width=0.4\textwidth]{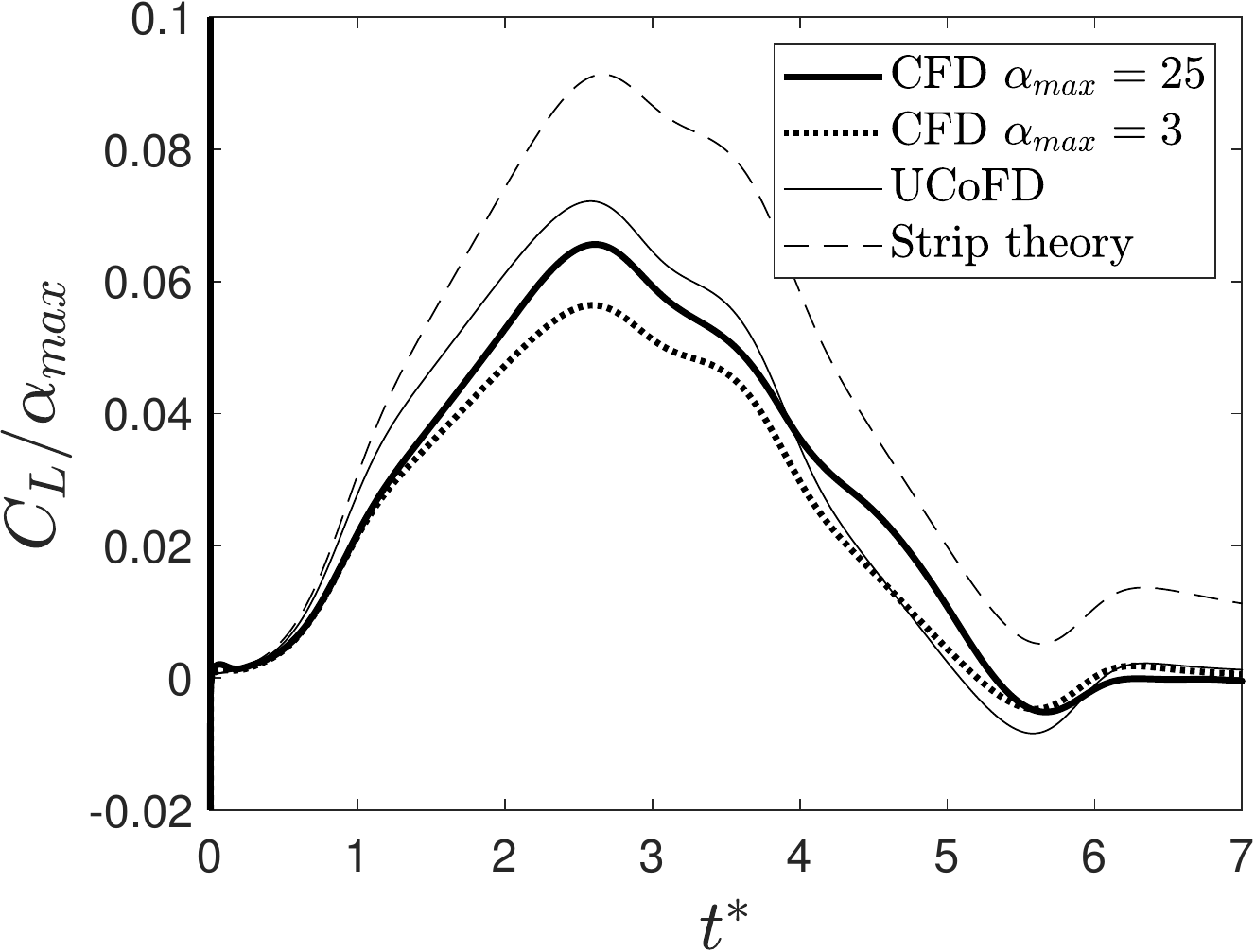}}
    \subfigure[Aspect ratio 3 $C_M$]{
	\label{fig:eldredge_pitch25_ar3_CM}
    	\includegraphics[width=0.4\textwidth]{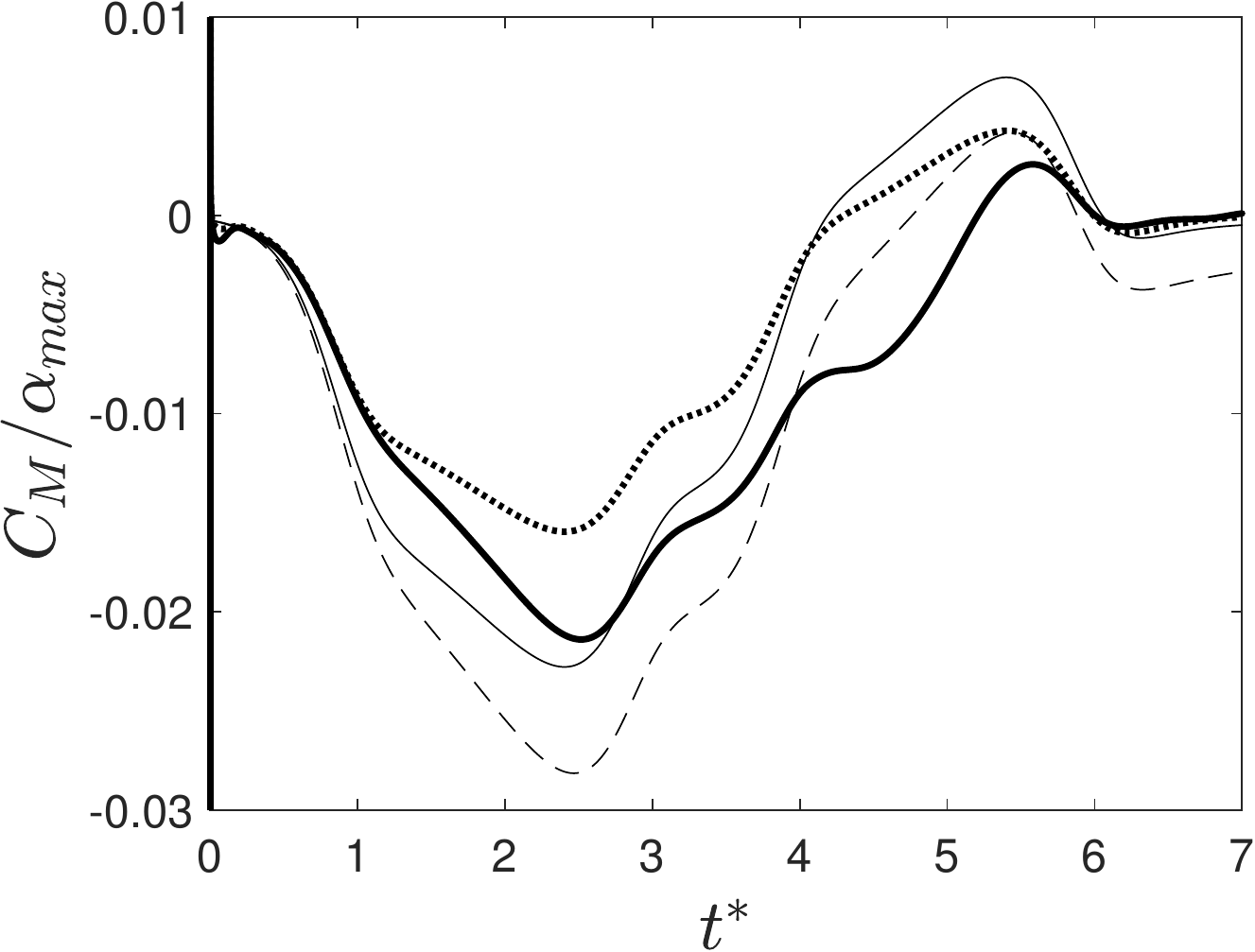}}

	\caption{Comparison of UCoFD, CFD and strip theory results for aspect ratio 6 and 3 rectangular wings undergoing
	a smooth pitch ramp maneuver to angles of attack of \ang{25} and \ang{3} at a Reynolds number of \num{10000}. }
	\label{fig:eldredge_pitch25_CM_CL}	
\end{figure}

The resulting lift and moment coefficients are shown in Fig.~\ref{fig:eldredge_pitch25_CM_CL}.
The coefficients are normalized by the maximum angle of attack to allow for comparison. 
Both UCoFD and strip theory are linear, consequently their results are independent of the amplitude of the kinematics. The
differences between the $\alpha_{\text{max}}= \ang{25}$ and $\alpha_{\text{max}}= \ang{3}$ CFD results
are due to aerodynamic non-linearities.

The results for the aspect ratio 6 cases, shown in Fig.~\ref{fig:eldredge_pitch25_ar6_CL} and  
Fig.~\ref{fig:eldredge_pitch25_ar6_CM} will be considered first. 
The assumption made in the derivation on lifting-line theory that the aspect ratio is high is
better satisfied by this \AR 6 wing than the \AR 4 wing used earlier. 
For the small amplitude $\alpha_{\text{max}}= \ang{3}$, the UCoFD method predicts the shape of both
the $C_L$ and $C_M$ curves excellently. However, it overestimates the peak
lift and moment compared to the CFD, and by a much greater margin than 
for the Euler cases studied earlier. This is due to neglecting viscous effects which are present in the CFD at Re$=10000$.
However, the overestimation is much smaller than that produced by strip theory.

\begin{figure}
\centering
\begin{tabular}{ >{\centering\arraybackslash}m{0.05\textwidth}  >{\centering\arraybackslash}m{0.25\textwidth}   >{\centering\arraybackslash}m{0.25\textwidth}  >{\centering\arraybackslash}m{0.25\textwidth}}
\toprule
&$t^* = 3.0$ & $t^* = 4.5$ & $t^* = 5.5$\\
\midrule
	\AR 6
	&\includegraphics[width=0.25\textwidth,trim={16cm 6cm 8cm 1cm},clip]{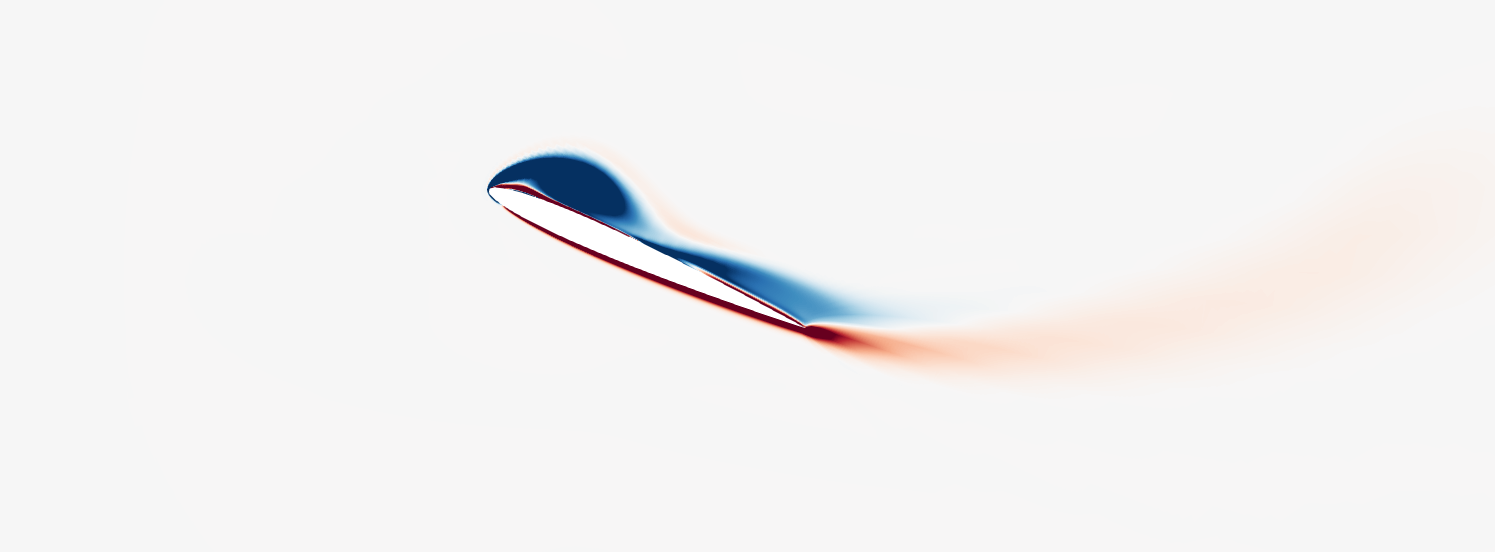}
	&\includegraphics[width=0.25\textwidth,trim={16cm 6cm 8cm 1cm},clip]{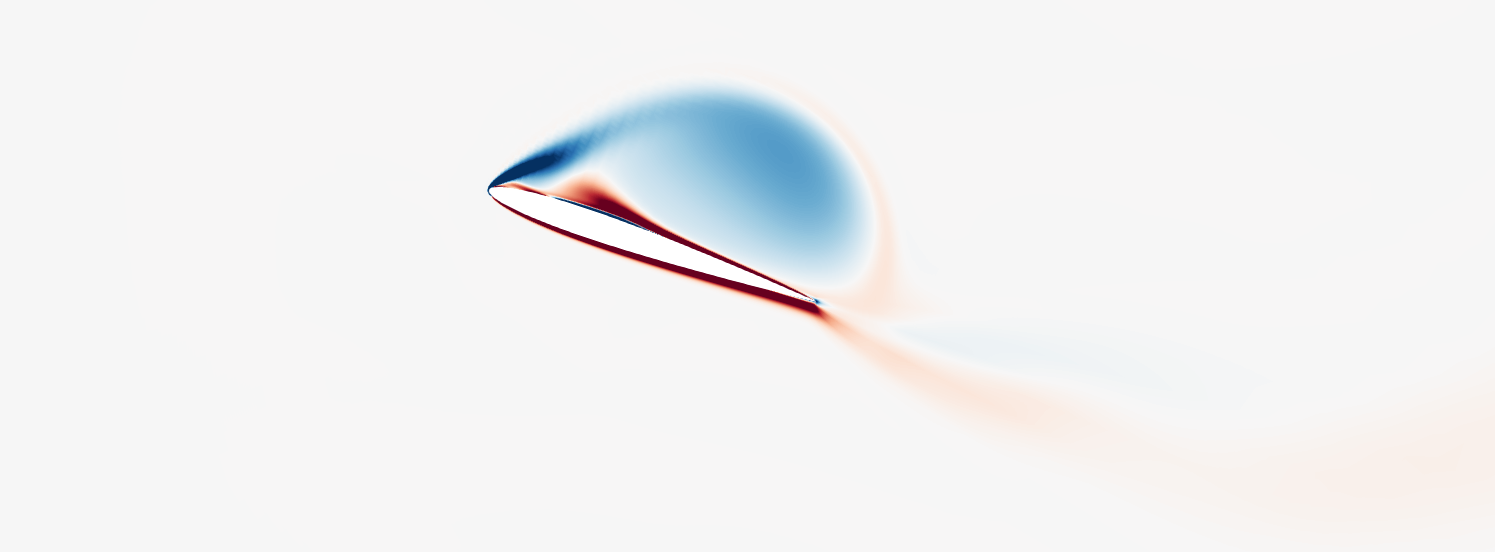}
	&\includegraphics[width=0.25\textwidth,trim={16cm 6cm 8cm 1cm},clip]{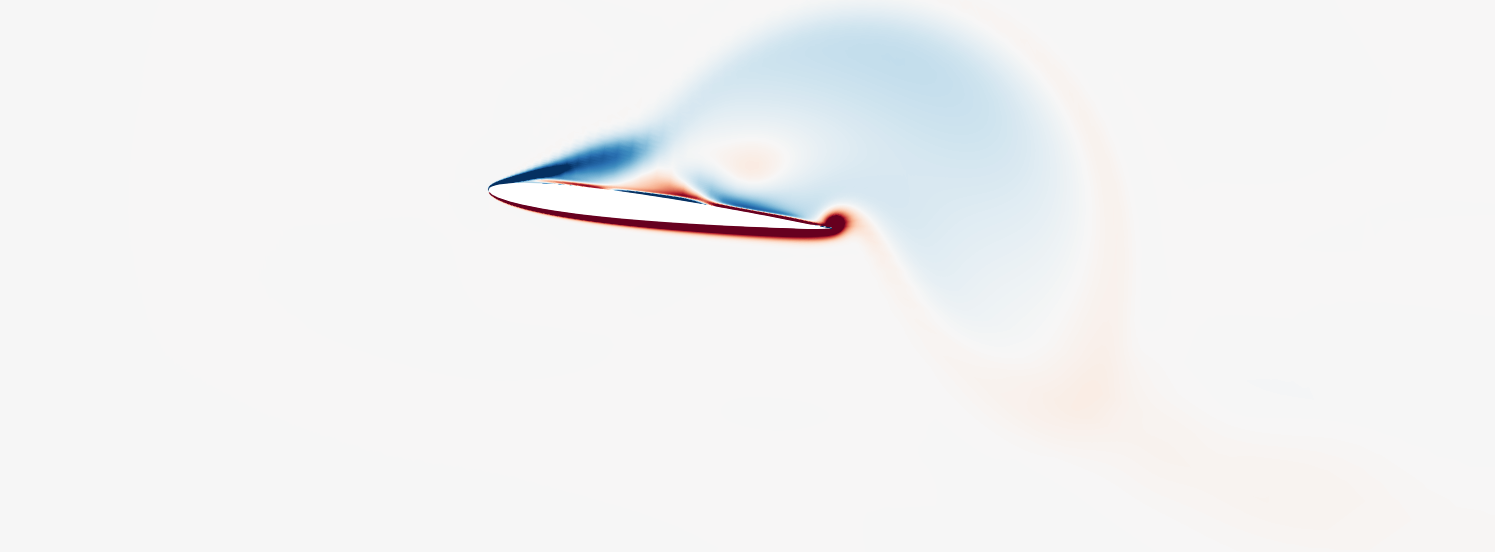}\\
	\AR3
	&\includegraphics[width=0.25\textwidth,trim={16cm 6cm 8cm 1cm},clip]{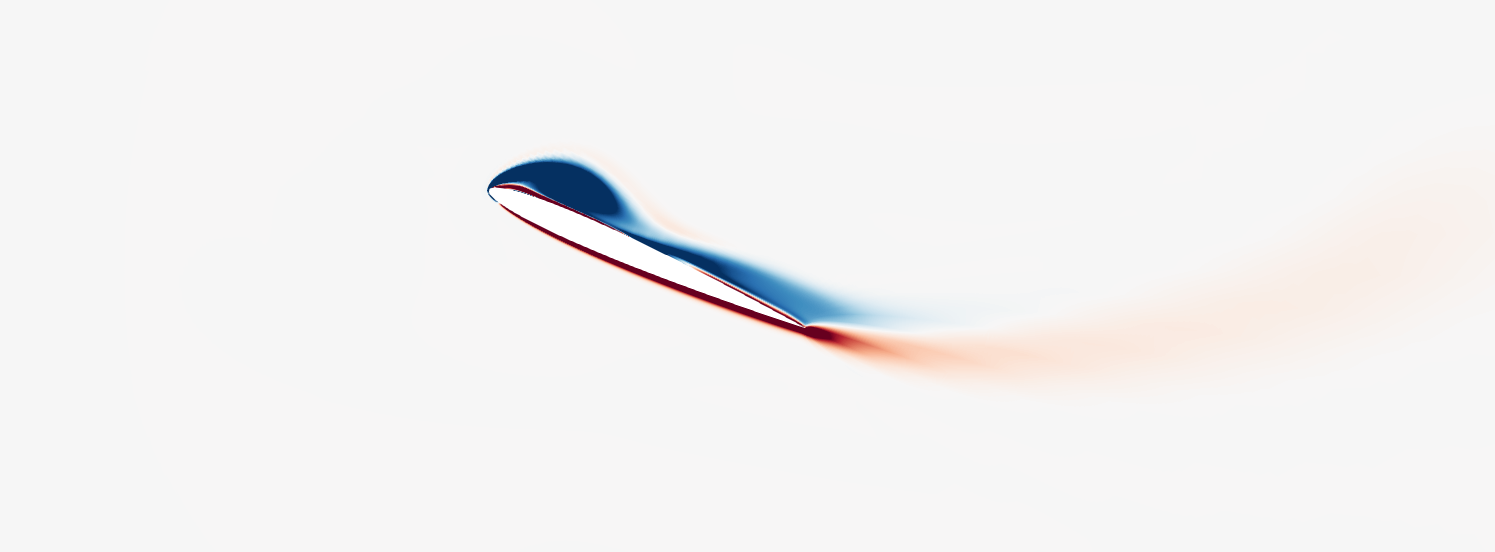}
	&\includegraphics[width=0.25\textwidth,trim={16cm 6cm 8cm 1cm},clip]{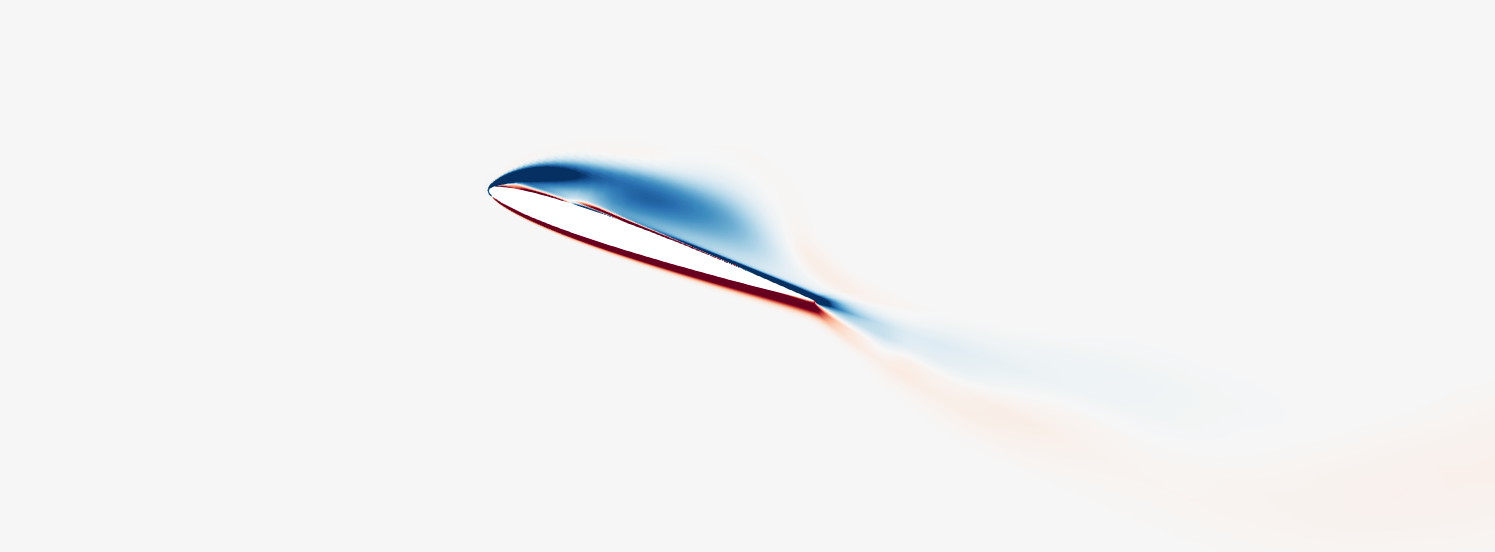}
	&\includegraphics[width=0.25\textwidth,trim={16cm 6cm 8cm 1cm},clip]{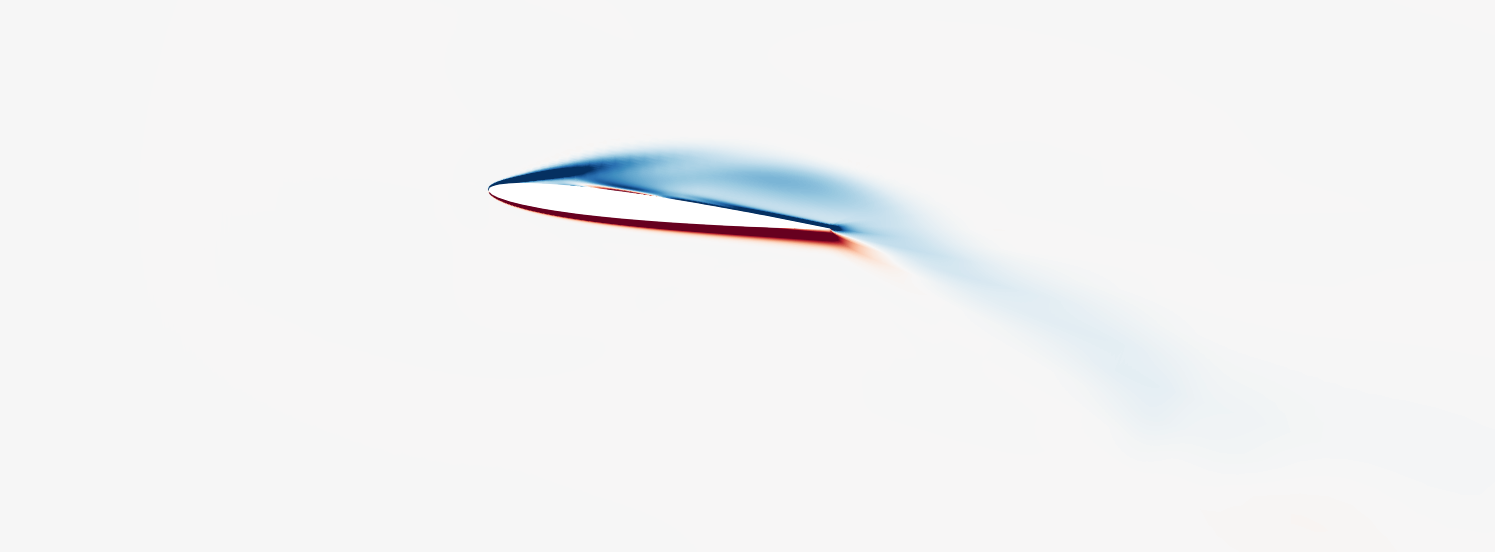}\\
	    	
\bottomrule
\end{tabular}
	\includegraphics[width=0.65\textwidth]{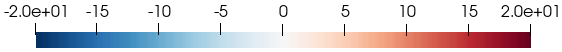}
		\caption{Non-dimensionalized spanwise vorticity $\omega_y c/U$ at the root section of aspect ratio 6 and 3 wings, obtained from CFD
		at a Reynolds number of \num{10000}. }
		\label{fig:big_amp_pitch_vort}
\end{figure}

Increasing the amplitude of the kinematics from  $\alpha_{\text{max}}= \ang{3}$ to
 $\alpha_{\text{max}}= \ang{25}$ has no impact on UCoFD results since the theory is linear.
However, the difference in the CFD results is significant. The peak forces increase super-linearly,
and the shape of the $C_L$ and $C_M$ curves change significantly, particularly between
$t^*=4$ and $t^*=5$. This is due to the formation of a leading-edge vortex, that then detaches
from the leading edge before being convected over the upper surface of the wing. 
This process is shown for both wings in Fig.~\ref{fig:big_amp_pitch_vort}. The UCoFD
cannot model this aerodynamic non-linearity. Consequently, it is unable to reflect the 
change in shape of the $C_L$ and $C_M$ curves obtained from CFD. However, since
the CFD result predicted super-linearly increasing forces, the overestimate of forces
by the UCoFD has reduced. The strip theory result continues to overestimate forces 
more significantly than UCoFD.

At aspect ratio 3, the UCoFD over-prediction of forces in comparison to the CFD results
is greater. Again, for the $\alpha_{\text{max}}= \ang{3}$ cases, the UCoFD method correctly
predicts the CFD result curve shapes. For the large amplitude $\alpha_{\text{max}}= \ang{25}$ case,
the CFD force curves are more similar to the $\alpha_{\text{max}}= \ang{3}$ case than at 
aspect ratio 6. The lower aspect ratio results in a smaller and more stable leading edge-vortex (seen in Fig.~\ref{fig:big_amp_pitch_vort}). Consequently, the
UCoFD prediction of the $\alpha_{\text{max}}= \ang{25}$ CFD curves
is better, but not excellent. In comparison, the strip theory results become worse with 
decreasing aspect ratio due to the increased importance of finite-wing effects.

\section{Conclusions}
\label{sec:conclusions}

The Unsteady Lifting-Line Theory / Convolution in Frequency Domain (UCoFD)
method was derived: it allows frequency-domain unsteady lifting-line theory to be applied to time domain
problems by taking the Fourier transform of the input kinematics, performing
convolution with interpolated frequency-domain unsteady lifting-line results,
and returning the solution to the time domain via the inverse Fourier transform.

Quadratic interpolation was used to obtain frequency domain results from 
several evaluations of an unsteady lifting-line theory. The Fourier transform of the time-domain kinematics, 
obtained using the fast Fourier transform method, is then convolved with with this interpolation, before 
the inverse Fourier transform is applied to obtain time-domain results. Since
the frequency-domain results are independent of the time-domain kinematics, they
need only be computed once for a given wing geometry.

This method was compared to CFD results for a rectangular aspect ratio 4 wing
in the Euler regime. Excellent results were obtained for both a pitch ramp case,
and a heave velocity ramp case. The heave velocity ramp case also demonstrated
how kinematics with different start and end points can be simulated with the current method.
Compared to strip theory, the UCoFD method gave much better results.

Finally, the method was compared to a low-Reynolds number, large amplitude case.
Here, the underlying assumptions used in the derivation of the method were broken,
since the method doesn't model the leading edge vortex found in the CFD
results. Nonetheless, UCoFD gave a reasonable prediction of lift and moment coefficients.

\section*{Acknowledgments}
The authors gratefully acknowledges the support of the UK
Engineering and Physical Sciences Research Council (EPSRC)
through a DTA scholarship, grant EP/R008035 and the
 Cirrus UK National Tier-2 HPC service at 
 EPCC (http://www.cirrus.ac.uk).

\bibliography{JabRefDatabase,kiran_bibtot}
\end{document}